\documentclass[preprint,showpacs,superscriptaddress]{revtex4-1}
\usepackage{graphicx}
\usepackage{dcolumn}
\usepackage{bm}
\usepackage{longtable}
\usepackage[usenames]{color}
\usepackage{pstricks} 
\begin{document}

\title{
 Quantum dynamics of CO-H$_2$ in full dimensionality
}
\author{Benhui Yang}
\affiliation{Department of Physics and Astronomy and the Center for 
  Simulational Physics, The University of Georgia, Athens, GA 30602, USA}
\author{P. Zhang}
\affiliation{Department of Chemistry, Duke University, Durham, NC 27708, USA}
\author{X. Wang}
\affiliation{Department of Chemistry, Emory University, Atlanta, GA 30322, USA}
\author{P. C. Stancil}
\email{stancil@physast.uga.edu}
\affiliation{Department of Physics and Astronomy and the Center for 
  Simulational Physics, The University of Georgia, Athens, GA 30602, USA}
\author{J. M. Bowman}
\affiliation{Department of Chemistry, Emory University, Atlanta, GA 30322, USA}
\author{N. Balakrishnan}
\affiliation{Department of Chemistry, University of Nevada Las Vegas, Las Vegas,
  NV 89154, USA}
\author{R. C. Forrey}
\affiliation{Department of Physics, Penn State University,
   Berks Campus, Reading, PA 19610, USA}

\date{\today}

\begin{abstract}
Accurate rate coefficients for molecular vibrational transitions due
to collisions with H$_2$, critical for interpreting infrared astronomical observations, 
are lacking for most molecules. Quantum calculations are the primary source of such 
data, but reliable values that consider all internal degrees of 
freedom of the collision complex have
only been reported for H$_2$-H$_2$ due to the difficulty of the computations. 
Here we present essentially exact full-dimensional dynamics computations for rovibrational
quenching of CO due to H$_2$ impact. Using a high-level
six-dimensional potential surface, time-independent scattering calculations,
within a full angular-momentum-coupling formulation, 
were performed for the deexcitation of 
vibrationally excited CO. Agreement with 
experimentally-determined results confirms the accuracy of the potential
and scattering computations, representing the largest of such calculations
performed to date. This investigation advances 
computational quantum dynamics studies representing
initial steps toward  obtaining CO-H$_2$ rovibrational quenching data 
needed for astrophysical modeling.
\end{abstract}


\maketitle


Quantum mechanical studies of inelastic processes in molecular collisions 
began with the development of the nearly-exact, close-coupling (CC) method for 
rotational transitions in atom-diatom
collisions by Arthurs and Dalgarno \cite{art60} and Takayanagi \cite{tak65} 
in the 1960s. Over the past five decades tremendous advances 
in computational processing power and numerical algorithms
have allowed high-level computations of inelastic processes
\cite{dub13,rou13} as well as  
reactive collisional
dynamics of large molecular systems, the latter using primarily time-dependent
approaches \cite{alt03,bha10,bha12}. 
Until recently, however, the largest full-dimensional inelastic studies 
for a system of two colliding  molecules have been limited to H$_2$-H$_2$ 
collisions in six-dimensions (6D),
which were performed with both time-independent \cite{pog02} and time-dependent \cite{lin03,pan07} 
approaches for solving the Schr\"odinger equation.
However, to make these computations possible, the authors resorted to
various angular-momentum decoupling approximations with uncertain reliability. It was only
recently~\cite{que08,que09,sam11} that 
these decoupling approximations were relaxed and essentially
exact full-dimensional CC computations for the H$_2$-H$_2$ system became feasible.
Despite the large internal energy spacing of H$_2$, which allows the basis sets to be
relatively compact, these calculations are computationally demanding. 
Replacing one H$_2$ molecule with a molecule that has smaller internal energy spacing,
such as carbon monoxide,    
would further increase the computational demands.  Whether the CC method could 
in practice be used to describe such a diatom-diatom system in full-dimensionality
 remains an open question.

First detected
in the interstellar medium in 1970 \cite{wil70}, carbon monoxide is the second most abundant molecule,
after H$_2$, in most astrophysical environments.
CO has been the focus of countless theoretical astrophysical studies and observations, being detected 
in objects as distant as high redshift quasars \cite{nar08} to 
cometary comae in our solar system \cite{lup07} to the atmospheres of extrasolar giant planets
\cite{swa09}. Most studies have focused on pure rotational
transitions observed in the far infrared to the radio or electronic absorption in the near ultraviolet. Over the past decade, however,
near infrared (NIR) emission of CO due to the fundamental vibrational
band near 4.7 $\mu$m has been detected in a variety of sources including star-forming regions in Orion with the {\it Infrared
Space Observatory} \cite{gon02} and protoplanetary
disks of young stellar objects \cite{bri09,bro13,hei14} with the Gemini Observatory and the Very Large Telescope. 
In addition, pure rotational transitions, but in the first
two vibrationally excited states ($v_1=1$ and 2), were detected by the Submillimeter Array in the
circumstellar shell of the much-studied evolved star IRC+10216 \cite{pat09}.
In particular, high resolution observations of the CO fundamental band probe the physical conditions
in the inner disk region, $\sim$10-20 astronomical units (AU), the site of planet formation.
Detailed modeling of such environments requires state-to-state inelastic
rovibrational excitation rate coefficients for CO due to H$_2$ collisions, but current simulations
are limited to approximate scaling methods due to the
dearth of explicit data \cite{gon02,thi13}.

In this Article, we address the two issues outlined above by advancing the state-of-the-art 
for inelastic quantum dynamics with a
full-dimensional investigation using an accurate potential energy surface relevant 
to this scattering process, with a particular emphasis on the important region of the 
van der Waals complex. This
was made possible through the accurate computation and precise fitting of a 6D CO-H$_2$
potential energy surface (PES) 
in the relevant region
of the formaldehyde tetra-atomic system and the further
development of the  inelastic diatom-diatom time-independent scattering code, TwoBC \cite{rom06},
which performs full angular-momentum coupling, the 
CC formalism \cite{art60}, including vibrational degrees-of-freedom. 
We first briefly
describe the new CO-H$_2$ PES and its testing through comparison of rotational excitation 
calculations using the 6D PES and  4D PESs (neglecting vibrational motion) and available
experimental results. The full-dimensional (6D), essentially exact, computations for
rovibrational quenching of CO($v_1=1$) due to H$_2$ collisions are presented resolving
a two-orders-of-magnitude discrepancy  between earlier 4D calculations which adopted various approximations \cite{bac85a,bac85b,rei97,flo12}. Finally, the current results 
are consistent with the  rovibrational
quenching measurements for the CO-H$_2$ system, performed at the Oxford Physical
Chemistry Laboratory
from 1976-1993, which no prior calculation has yet been able to adequately
explain \cite{and75,and76,wil93}.

\section*{Results}

{\bf The CO-H$_2$ potential energy surface.}
The CO-H$_2$ interaction has been of considerable interest to the chemical physics community
for many decades
with one of the first 4D surfaces for the electronic ground state constructed by Schinke {\it et al.} \cite{sch84}, which was later
extended by Ba\^ci\'c {\it et al.} \cite{bac85a}. The group of Jankowski, Szalewicz, and coworkers performed
accurate 5D and 6D electronic energy calculations, but averaged over monomer
vibrational modes, and performed several fits to obtain a series of 4D rigid-rotor surfaces,  referred to as the  V98 \cite{jan98}, V04~\cite{jan05}, and V12~\cite{jan13} PESs. A 6D PES
for formaldehyde was constructed earlier by Zhang {\it et al.}
\cite{zha04}, but it was developed for reactive scattering applications and 
consequently, limited attention was given to the long-range CO-H$_2$ 
van der Waals configuration. Therefore as a prerequisite to 6D inelastic dynamics studies, we
carried out an unprecedented  potential energy calculation including over 459,756 energy
points (see Methods for details). The potential energy data were then fit using the invariant polynomial
method with Morse-type variables in terms of bond-distances \cite{bow11,bra09}.
The resulting 6D PES, referred to as V6D, is shown in Fig.~\ref{surface} for one
sample configuration.  Some features of V6D are also illustrated in
Supplementary Figs.~S2 - S4.

{\bf Cross sections and rate coefficients.}
Time-independent quantum scattering calculations were performed using the
 CC formulation of Arthurs and Dalgarno \cite{art60} as implemented for diatom-diatom
 collisions in the 4D rigid-rotor approximation in MOLSCAT \cite{molscat} and   extended to
 full-dimensional dynamics as described by  Qu\'em\'ener, Balakrishnan,
 and coworkers \cite{que09,sam11} in TwoBC.
 In the first set of scattering calculations, the new 6D PES was tested for
 pure rotational excitation from CO($v_1=0,j_1=0,1$), where $v_1$ and $j_1$ are the 
 vibrational and rotational
 quantum numbers, respectively. The crossed
 molecular beam experiment of Antonova {\it et al.} \cite{ant00}, who obtained relative
 state-to-state rotational inelastic cross sections, is used as a benchmark.
 The experimental cross sections were determined at three center-of-mass kinetic energies
 (795, 860, and 991 cm$^{-1}$), but with an initial state distribution of CO
 estimated to be $75\pm5$\%  for $j_1=0$ and $25\pm5$\%  for $j_1=1$.  Antonova {\it et al.}
 normalized the relative cross sections to rigid-rotor calculations done with MOLSCAT
 using the V04 PES of Jankowski and Szalewicz \cite{jan98}. Comparison of
 the experiment with new 4D rigid-rotor calculations on V12 and full-dimensional 
calculations on V6D are shown in Fig.~\ref{ant-exp}.
  We find no difference in the excitation cross
 sections when using V04 or V12, while the RMS cross section errors between the 
 normalized experiment and the V12 and V6D calculations range from
 0.56-0.89$\times 10^{-16}$ cm$^2$ and 0.55-0.95$\times 10^{-16}$ cm$^{2}$, respectively
 (See Supplementary Table S1 and Supplementary Note 1).
Clearly, a 4D rigid-rotor treatment of the dynamics is sufficient for describing rotational
excitation at these relatively high energies.

The importance of full dimensionality for rotational excitation
becomes more evident as the collision energy is reduced
(see also Supplementary Figs. S5 and S6). 
Low-energy excitation cross sections for the process
\begin{equation}
{\rm CO}(v_1=0,j_1=0) + {\rm H}_2(v_2=0,j_2=0) \rightarrow
 {\rm CO}(v_1^\prime=0,j_1^\prime=1) + {\rm H}_2(v_2^\prime=0,j_2^\prime=0),
\label{eqrot}
\end{equation}
or (0000)$\rightarrow$(0100), using the notation defined in Methods,
were measured by Chefdeville {\it et al.} \cite{che12} in a crossed-beam
experiment.  They obtained the excitation cross section for center of mass
kinetic energies from 3.3 to 22.5 cm$^{-1}$. Though their
energy resolution was limited, three broad features were detected
near 6, 13, and 16 cm$^{-1}$ attributed to orbiting
resonances. Computed cross sections for process~(\ref{eqrot})
using the 4D V12 PES  and the full-dimensional V6D PES are presented 
in Fig.~\ref{gau-rigid}a. While both computations reveal numerous resonances,
the resonances are generally shifted by 2-3 cm$^{-1}$ between calculations. The energy and magnitude 
of the resonances are very sensitive to the details
of the PESs, but differences may also be due to the relaxation of the rigid-rotor
approximation with the use of V6D in TwoBC. In Fig.~3b, the two calculations
are convolved over the experimental energy resolution and compared
to the measured relative cross sections. Except for
the peak near 8 cm$^{-1}$, the current 6D calculation appears to reproduce the
main features of the experiment. RMS errors are found to be 0.355 and
0.228$\times 10^{-16}$ cm$^2$ for the V12 and V6D PESs, respectively.
Further details on the rotational excitation calculations can be found in 
Supplementary Note 1.  

Now that the improved performance of the V6D potential for pure
rotational excitation is apparent, we turn to rovibrational transitions. As
far as we are aware, prior experimental \cite{and75,and76,wil93,hoo63,mil64}  and theoretical \cite{bac85a,bac85b,rei97,flo12} studies
are limited to the total quenching of CO($v_1=1$). In the scattering calculations of
both Ba\^ci\'c {\it et al.} \cite{bac85a,bac85b} and Reid {\it et al.} \cite{rei97}, the
4D potential of Ba\^ci\'c {\it et al.} \cite{bac85a} was adopted, in which two
coordinates  were fixed ($r_2=1.4$~a$_0$, $\phi=0$), and various combinations of 
angular-momentum decoupling approximations for the dynamics were utilized
(e.g., the infinite order sudden, IOS, and coupled-states, CS, approximations; 
see Supplementary Note 2).
More recently, Flower \cite{flo12} performed
CC calculations on a parameterization of the 4D PES of Kobaysashi {\it et al.}
\cite{kob00}. These four sets of 4D calculations for quenching due to
para-H$_2$ ($j_2=0$) are compared in Fig.~\ref{total-cross} to the
current 6D/CC calculations on the V6D surface for the case of $j_2=j^\prime_2=0$. 
Corresponding state-to-state and total cross sections
for collisions with ortho-H$_2$ are given in 
Supplementary Figs. S7-S9 with further details in Supplementary Note 2

A cursory glance at Fig.~\ref{total-cross}  reveals a
more than two orders of magnitude discrepancy among the various
calculations.  The large dispersion 
for the previous calculations is due to a combination of reduced 
dimensionality and decoupled angular momentum which makes it 
difficult to assess the reliability of each approximation.
The current results, however, remove these uncertainties by  
utilizing  i) a full-dimensional (6D) PES, ii) full-dimensional (6D) dynamics,
and iii) full angular-momentum coupling. The sharp peaks in the cross sections over the 
1-10 cm$^{-1}$ range in the 6D/CC results are due to resonances 
\cite{cha10,cas13} supported by the CO-H$_2$ van der Waals 
potential well. These resonances, in the rovibrational quenching of CO by H$_2$, are reported here for the first
time and their prediction is made possible through our high-level treatment of the dynamics. Computations for transitions involving other
initially excited $j_2$ states and inelastic H$_2$ ($j_2\neq j_2^\prime$) transitions are
presented and compared in Supplementary Figs. S10 and S11 and Supplementary Note 2.

Using the current 6D/CC cross sections and the 4D/IOS-CS results of Reid {\it et al.},
rate coefficients for a Maxwellian velocity distribution are computed and compared
in Fig.~\ref{total-rate}
to total deexcitation measurements \cite{and75,and76,wil93} reported by
Reid {\it et al.} \cite{rei97}. 
The comparison is not straight-forward because i) the
measurements correspond to an initial thermal population of H$_2$ rotational states,
ii) the initial rotational population of CO($v_1=1,j_1$) was unknown, iii) the
experimental rate coefficients for ortho-H$_2$ are estimated from para- and normal-H$_2$
measurements, and iv) the contribution from a quasi-resonant channel,
\begin{equation}
{\rm CO}(v_1=1) + {\rm H}_2(v_2=0,j_2=2) \rightarrow
 {\rm CO}(v_1^{\prime}=0) + {\rm H}_2(v_2^{\prime}=0,j_2^{\prime}=6) + \Delta E ,
\label{qr}
\end{equation}
dominates the para-H$_2$ case for
$T\gtrsim 50$~K. 
Fig.~\ref{total-rate}a displays the CO($v_1=1$) rovibrational deexcitation
rate coefficients from the current 6D/CC calculations for collisions of ortho-H$_2$ for $j_2$=1 and 3, separately. These rate
coefficients are summed over all $j_1^\prime$ and $j_2^\prime$=1, 3, and 5. Assuming
a Boltzmann average at the kinetic temperature $T$ of the H$_2$ rotational levels $j_2$ = 1 and 3, presumed 
to be representative of the experimental conditions, rate coefficients are computed and found to be
in good agreement with the measurements above 200~K. From Fig.~\ref{total-rate}a, the 
contribution from $j_2$=3 is seen to be only important above 150 K. The remaining
difference with experiment at low temperatures may be due to the fact mentioned above that the
ortho-H$_2$ rate coefficients are not directly measured, but deduced from normal-H$_2$ and para-H$_2$
experiments. In particular, Reid {\it et al.} assume that the ortho/para ratio in the normal-H$_2$ measurements
is 3:1, i.e. statistical, and independent of temperature.
Further, as stated above, the experimental CO rotational population distribution in $v_1=1$ is unknown. Nevertheless,
the current 6D/CC computations are a significant advance over the 4D results of Reid {\it et al.}
which also correspond to a Boltzmann average of rate coefficients for $j_2$=1 and 3.

As indicated above, the situation for para-H$_2$ collisions is more complicated due to the
quasi-resonant contribution (\ref{qr}), a mechanism not important for ortho-H$_2$. Boltzmann-averaged
rate coefficients are presented in Fig.~\ref{total-rate}b including $j_2$=0 and 2 summed over
$j_2^\prime$= 0, 2, and 4, with and without the quasi-resonant contribution, $j_2=2 \rightarrow j_2^\prime =6$.
While the current 6D/CC results and the 4D calculations of Reid {\it et al.} are in agreement that the quasi-resonant
contribution becomes important for $T\gtrsim 50$~K, the relative magnitude compared to the non-resonant
transitions from the 6D/CC calculation is somewhat less than obtained previously with the 4D potential.
This is partly related to the fact that the 6D/CC rate coefficients for $j_2$=0 are significantly larger
than those of Reid {\it et al.} (see also the corresponding cross sections in Fig.~\ref{total-cross}
and in the Supplementary Fig. S9).
Compared to the experiment, we obtain excellent
agreement for $T\lesssim 150$~K, but are somewhat smaller at higher temperatures. This small discrepancy 
may be related to the unknown CO($v_1=1,j_1$) rotational population in the measurement.
Nevertheless, it is only the 6D/CC computations, i.e., dynamics in full dimensionality with full angular
momentum coupling, which can reproduce the measurements for both ortho- and para-H$_2$.
Further, the computations for the quasi-resonant process (2) were the most challenging reported
here due to the requirement of a very large basis set (see Supplementary Table S2) 
which resulted in long computation
times, a large number of channels, and usage of significant disk space ($\sim$0.5 TB per partial wave). 
In total, the cross sections given here  consumed $>$40,000 CPU hours.

\section*{Discussion}

The current investigation of the CO-H$_2$ inelastic collision system has been performed with the
intent of minimal approximation through the computation of a high-level potential energy surface,
robust surface fitting, and full-dimensional inelastic dynamics with full angular-momentum coupling.
That is, within this paradigm for studying inelastic dynamics, we have advanced the state-of-the-art
for diatom-diatom collisions through this unprecedented series of computations.
The approach has been benchmarked against experiment for pure rotational and rovibrational
transitions giving the most accurate results to date within the experimental uncertainties and
unknowns. The accuracy and long-range behavior of the 6D potential energy surface is found to be comparable to previous,
lower-dimensional surfaces. The agreement of the current computation for the CO($v_1=1$)
rovibrational quenching with measurement
resolves a long-standing (more than two decades) discrepancy and
justifies the requirement of a full-dimensional approach. This methodology can now, though
with significant computational cost, be applied to a large range of initial rotational levels for $v_1=1$
and for higher vibrational excitation to compute detailed state-to-state cross sections unobtainable
via experiment.

This advance in computational inelastic scattering is particularly timely as ground-based (e.g., the Very
Large Telescope (VLT))
observations have focused on CO rovibrational emission/absorption in a variety of
astrophysical objects, while related observations are in the planning stages for future
space-based telescopes (e.g., NASA's {\it James Webb Space Telescope}). In particular,  we are in an exciting era of investigation into the
properties of protoplanetary disks (PPDs) around young stellar objects \cite{hen13}. PPDs provide the
material for newly-forming stars and fledgling planets. The
CO fundamental band ($|\Delta v_1| =1$) is a tracer of warm gas in the inner regions of PPDs
and, with appropriate modeling, gives insight into disk-gas kinematics and disk evolution in
that zone where planets are expected to be forming.
A recent survey of 69 PPDs with the VLT \cite{bro13} detected CO vibrational bands in 77\% of the sources
including $v_1=1\rightarrow 0$, $v_1=2\rightarrow 1$, $v_1=3\rightarrow 2$, and even $v_1=4\rightarrow 3$
in a few cases. Remarkably, rotational excitation as high as $j_1=32$ was observed. However, the
modeling of PPDs, and other astrophysical sources with CO vibrational excitation, is  hindered
by the lack of rate coefficients due to H$_2$ collisions. We are now in an excellent
position to provide full-dimensional, state-to-state CO-H$_2$ collisional data which will not only have a profound impact on models
characterizing these  intriguing environments that give birth to planets, but also aid in critiquing current theories used to
describe their evolution.

\section*{Methods}

{\bf  Potential energy computations.}
The potential energy computations were performed using the explicitly correlated
coupled-cluster (CCSD(T)-F12B) method \cite{ccsdf12, densityfit}, as
implemented in MOLPRO2010.1 \cite{molpro}.  
The cc-pcvqz-f12 orbital 
basis sets \cite{ccf12basis} that have been specifically optimized for use with 
explicitly correlated F12 methods and for core-valence correlation effects have been adopted. 
Density fitting approximations \cite{densityfit} were used in all explicitly correlated calculations 
with the AUG-CC-PVQZ/JKFIT and AUG-CC-PWCVQZ-MP auxiliary basis sets \cite{hat05,wei02}.
The diagonal, fixed amplitude 3C(FIX) ansatz was used, which is orbital invariant, 
size consistent, and free of geminal basis set superposition error (BSSE) \cite{tew06,fel10}.
The default CCSD-F12 correlation factor was employed in all calculations and 
all coupled cluster calculations assume a frozen core (C:1s and O:1s). 
The counter-poise (CP) correction \cite{cp}  was employed to reduce BSSE. Even though the explicitly 
correlated calculations recover a large fraction of the correlation energy, the CP  
correction is still necessary, mainly to reduce the BSSE of the Hartree-Fock contribution.
Benchmark calculations at the CCSD(T)-F12b/cc-pcvqz-f12 level were carried out 
on selected molecular configurations and results were compared 
with those from the conventional CCSD(T) method using aug-cc-pV5Z and aug-cc-pV6Z basis 
sets. Results showed that the CP corrected interaction energy agrees closely with those
derived from CCSD(T)/aug-cc-pV6Z. 

To construct the potential energy surface (PES), the computations were performed on a 
six-dimensional (6D) grid using Jacobi coordinates as shown in Supplementary Fig. S1.  
$R$ is the distance between the center-of-mass of CO and H$_2$. $r_1$ and $r_2$ are the bond lengths
of CO and H$_2$, respectively. $\theta_1$ is the angle between {\bf r}$_1$ and {\bf R},
$\theta_2$ the angle between {\bf r}$_2$ and {\bf R}, and $\phi$ 
the out-of-plane dihedral or twist angle.
In the potential energy computations, the bond lengths are taken over the ranges
$1.7359 \leq r_1 \leq 2.5359$~a$_0$  and $1.01 \leq r_2 \leq 1.81$~a$_0$,  
both with a step-size of 0.1~a$_0$.  For 
$R$, the grid extends from 4.0 to  18.0~a$_0$ with step-size of 0.5~a$_0$ for $R <$ 11.0~a$_0$ 
and 1.0~a$_0$ for $R >$ 11.0~a$_0$. All angular coordinates were computed with a step-size 
of 15$^\circ$ with
$0 \leq \theta_1 \leq 360^\circ$ and $0 \leq \theta_2,\phi \leq 180^\circ$. Additional points were
added in the region of the van der Waals minimum. 

\vskip 2.0cm

{\bf The PES fit.}
The CO-H$_2$ interaction PES has been fitted in 6D using an invariant polynomial
method \cite{bow11,bra09}. The PES was expanded in the form,
\begin{equation}
  V(y_1\cdots y_6)=\sum_{n1\cdots n6}^{N}c_{n_1\cdots n_6}y_1^{n_1}y_6^{n_6}
    [y_2^{n_2}y_3^{n_3}y_4^{n_4}y_5^{n_5}
   + y_2^{n_5}y_3^{n_4}y_4^{n_3}y_5^{n_2}],
\end{equation}
where $y_i=e^{-0.5d_i}$ is a Morse-type variable. The internuclear distances $d_i$ between two atoms
are defined as $d_1 = d_{\text{HH}^\prime}$,  $d_2 = d_{\text{OH}^\prime}$,
$d_3 = d_{\text{CH}^\prime}$,
 $d_4 = d_{\text{CH}}$,  $d_5 = d_{\text{OH}}$, and  $d_6 = d_{\text{CO}}$.
The total power of the polynomial, $N=n_1+n_2+n_3+n_4+n_5+n_6$, was restricted to 6.
Expansion coefficients $c_{n_1\cdots n_6}$ were obtained using weighted
least-squares fitting for potential energies up to 10,000 cm$^{-1}$.
The root-mean-square (RMS) error in the fit of the PES is 14.22 cm$^{-1}$ which
included 398,218 different geometries.
This RMS error can be compared to that of 277 cm$^{-1}$ for the 6D reactive surface of
Zhang {\it et al.} \cite{zha04}.
From the computed energy points, the global minimum of the total potential corresponds to the
collinear arrangement H-H-C-O ($\theta_1=0, \theta_2=0, \phi=0$) with
a depth of -85.937 cm$^{-1}$ at $R=8.0$ a$_0$ with $r_1$ and $r_2$ at their respectively 
equilibrium positions. This compares to the global minimum of the interaction
potential obtained by Jankowski {\it et al.} \cite{jan13}
from their fitted V12 PES averaged over $r_1$ and $r_2$: $R=7.911$ a$_0$ 
and -93.651 cm$^{-1}$. Note that this comparison is only suggestive as the global minimum in the total 
and interaction potentials coincide only when bond-lengths ($r_1$ and $r_2$) are 
fixed at their equilibrium values 
as illustrated in Fig. 1 and which is not the case for V12.

\vskip 2.0cm

{\bf Scattering theory and computational details.}
The quantum scattering theory for a collision of an $S$-state atom with a rigid-rotor was 
developed \cite{tak65,gre75,ale77,zar74} 
based on the close-coupling (CC) formulation of Arthurs and Dalgarno \cite{art60}.
Details about its extension to diatom-diatom collisions with full vibrational
motion can be found in Refs.~\cite{que09,sam11}. In this approach, 
the interaction potential $V(R,r_1,r_2,\theta_1,\theta_2,\phi)$ is expanded as,
\begin{equation}
V(R,r_1,r_2,\theta_1,\theta_2,\phi) = \sum^{}_{\lambda_1\lambda_2\lambda_{12}} 
A_{\lambda_1,\lambda_2,\lambda_{12}}(r_1,r_2,R)  Y_{\lambda_1,\lambda_2,\lambda_{12}}
(\hat{r}_1,\hat{r}_2,\hat{R}) ,
\label{v-lbd}
\end{equation}
with the bi-spherical harmonic function expressed as,
\begin{eqnarray}
\lefteqn{Y_{ \lambda_1,\lambda_2,\lambda_{12}}(\hat{r}_1,\hat{r}_2,\hat{R}) =
\sum^{}_{ m_{\lambda_1}m_{\lambda_2}m_{\lambda_{12}} } \big\langle  \lambda_1 m_{\lambda_1} \lambda_2 m_{\lambda_2} \big| 
 \lambda_{12} m_{\lambda_{12}} \big\rangle} \nonumber \\
 & & \times Y_{\lambda_1 m_{\lambda_1}}(\hat{r}_1) Y_{\lambda_2 m_{\lambda_2}}(\hat{r}_2) 
 Y^*_{\lambda_{12} m_{\lambda_{12}}}(\hat{R}) ,
\end{eqnarray}
where $ 0 \leq \lambda_1 \leq 10$, $0 \leq \lambda_2 \leq 6$ was used
in the scattering calculations. Due to the symmetry of H$_2$, only even values of
$\lambda_2$ contribute.

For convenience, the combined molecular state (CMS) notation is applied to describe a 
combination of rovibrational states for the two diatoms. 
A CMS represents a unique quantum state of the diatom - diatom system
before or after a collision. The CMS will be denoted as $(v_1j_1 v_2 j_2)$.
$v$ and $j$ are the vibrational and rotational quantum numbers.

The rovibrational state-to-state cross section as a function of collision energy $E$ is given by,
\begin{eqnarray}
\lefteqn{\sigma_{v_1j_1v_2j_2 \to v'_1j'_1v'_2j'_2}(E) = \frac{\pi}{(2j_1+1)(2j_2+1)k^2}} \nonumber  \\
&& \times \sum_{j_{12}j'_{12}ll'J\varepsilon_I}^{} (2J+1) 
|\delta_{v_1j_1v_2j_2l,v_1'j_1'v_2'j_2'l'} 
 - S^{J\varepsilon_I}_{v_1j_1v_2j_2l,v_1'j_1v_2'j_2'l'}(E)|^{2},
\end{eqnarray}
where ($v_1j_1v_2j_2$) and ($v_1^{\prime} j_1^{\prime}v_2^{\prime}j_2^{\prime}$) are, respectively, 
the initial and final CMSs of CO-H$_2$, 
the wave vector $k^2=2 \mu E/\hbar^2$, and $S$ is the scattering matrix. $l$ is the
orbital angular momentum and $J$ the total collision system angular momentum,
where {\bf J} = {\bf l} + {\bf j}$_{12}$ and {\bf j}$_{12}$ = {\bf j}$_1$ + {\bf j}$_2$.

Thorough convergence testing was performed in the scattering calculations
by varying all relevant parameters. The CC equations were propagated for
each value of $R$  from 4 to 18.0~a$_0$ 
using the log-derivative matrix propagation method of Johnson \cite{joh73}
and Manolopoulos \cite{man86}, which was found to converge for a  
radial step-size of $\Delta R=0.05$~a$_0$.
The convergence tests of the $v_1=1\rightarrow 0$ vibrational quenching cross section of CO 
with respect to the number of $v_1=1$ rotational channels found that at least
13-15 channels have to be included in the $v_1=1$ basis set, especially for low-energy scattering. 
Based on convergence tests with respect to the adopted maximum $R$ for the
long range part of the PES, we found that the cross sections are converged down to the
lowest collision energy of 0.1 cm$^{-1}$.  This value also guarantees that the rate coefficients
are converged for temperatures greater than 1~K. 
The number of discrete variable representation points $N_{r_1}$ and $N_{r_2}$;
the number of points in $\theta_1$ and $\theta_2$ for Gauss-Legendre quadrature,
$N_{\theta_1}$ and  $N_{\theta_2}$; and the number of points in $\phi$ for 
Gauss-Hermite quadrature, $N_{\phi}$, 
which were applied to project out the potential expansion coefficients were all 
tested for convergence with the final adopted values given in Supplementary Table~S2.
The basis sets and the maximum number of coupled channels are also presented 
in Supplementary Table~S2.

The resulting integral cross sections were 
thermally averaged over a Maxwellian kinetic energy distribution to yield state-to-state
rate coefficients as function of temperature $T$,  
\begin{equation}
k_{v_1j_1v_2j_2\rightarrow v_1^{\prime} j_1^{\prime}v_2^{\prime}j_2^{\prime}}(T) 
= \left (\frac{8}{\pi m \beta} \right )^{1/2}\beta^2\int^{\infty}_0 E 
\sigma_{v_1j_1v_2j_2\rightarrow v_1^{\prime} j_1^{\prime}v_2^{\prime}j_2^{\prime}}(E) 
\exp(-\beta E)dE,
\label{eq2}
\end{equation}
where $m$ is the reduced mass of the CO-H$_2$ complex,  
$\beta=(k_BT)^{-1}$, and $k_B$ is Boltzmann's constant.

\vskip 1.0cm
\noindent {\bf\large Acknowledgements}

Work at UGA and Emory was supported by NASA grant NNX12AF42G
from the Astronomy and Physics Research and Analysis Program,
at UNLV by NSF Grant No. PHY-1205838, 
and at Penn State by NSF Grant No. PHY-1203228.
This study was supported in part by resources and technical expertise from the
 UGA Georgia Advanced Computing Resource Center (GACRC), a partnership
 between the UGA Office of the Vice President for Research and Office of the
 Vice President for Information Technology. We thank  Shan-Ho Tsai (GACRC),
 Jeff Deroshia (UGA Department of Physics and Astronomy), and Gregg Derda
 (GACRC) for computational assistance.

\vskip 1.0cm
\noindent {\bf\large Author contributions}

BHY performed the potential energy surface (PES) fitting and scattering calculations.
PZ calculated the PES. XW and JMB developed the PES fitting code. NB, RCF, and PCS extended and 
modified the TwoBC code for CO-H$_2$ rovibrational scattering calculations, while NB assisted with the
scattering calculations. BHY, PCS, and PZ wrote
the article with contributions from all other authors.

\vskip 1.0cm
\noindent {\bf\large Additional information}

Supplementary information is included in the submission.

\vskip 1.0cm
\noindent {\bf\large Competing financial interests}

The authors declare no competing financial interests.

\newpage

\begin{figure}[hbt]
\includegraphics[scale=0.75,angle=0]{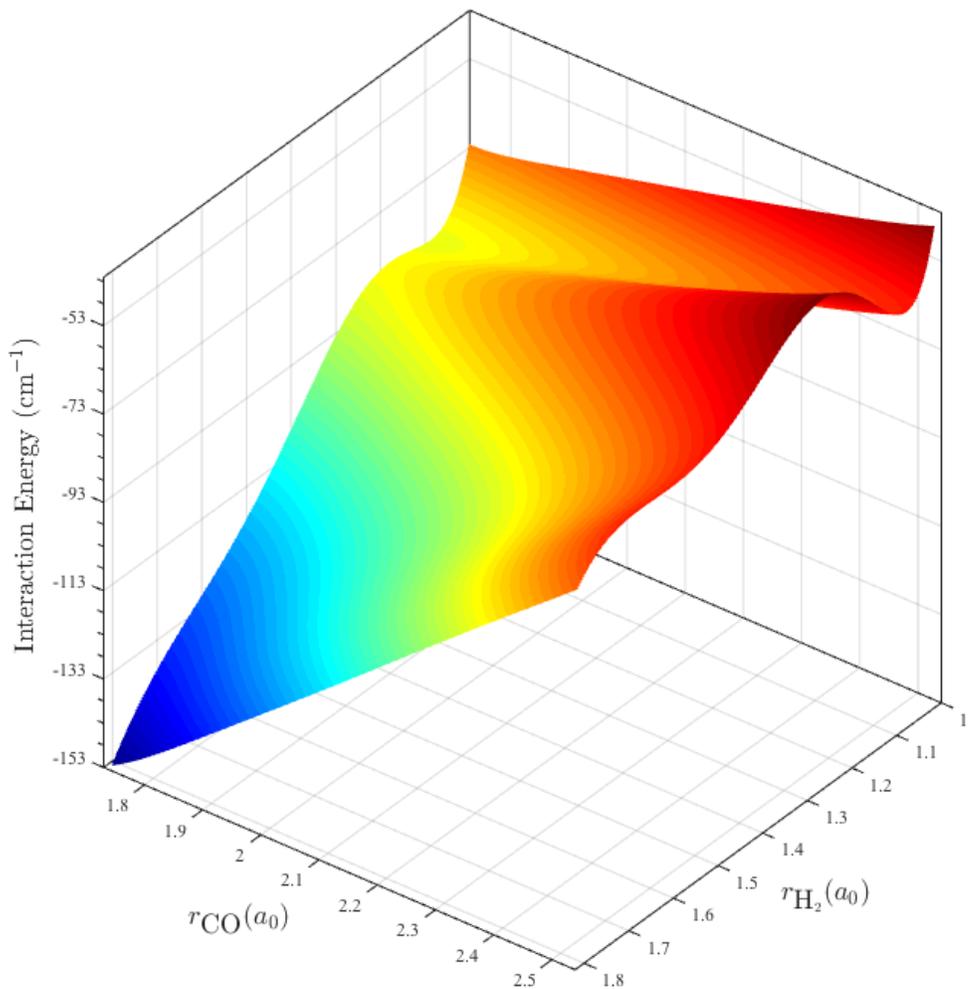}
\caption{{\bf The CO-H$_2$ interaction potential energy surface V6D.}
The potential surface is constructed in the 6D diatom-diatom Jacobi
coordinates ($R, r_1, r_2, \theta_1, \theta_2, \phi$),
$r_1$ and $r_2$ are bond lengths, $R$ the internuclear distance between CO and H$_2$
center of masses, $\theta_1$ and $\theta_2$ the angles between {\bf R} and {\bf r}$_1$ and {\bf r}$_2$,
and $\phi$ the dihedral or twist angle. See Supplemental Fig.~1.
Here the dependence of the potential
surface on bond lengths $r_{\rm CO} = r_1$ and $r_{\rm H_2}=r_2$ is shown with
$R=8$ a$_0$, $\theta_1=180^\circ$, $\theta_2$= 0, and $\phi$= 0.
Note that the CO($r_1$) and H$_2$($r_2$) diatom potentials have been subtracted.
}
\label{surface}
\end{figure}

\begin{figure}[hbt]
\advance\leftskip -0.0cm
\includegraphics[scale=0.9]{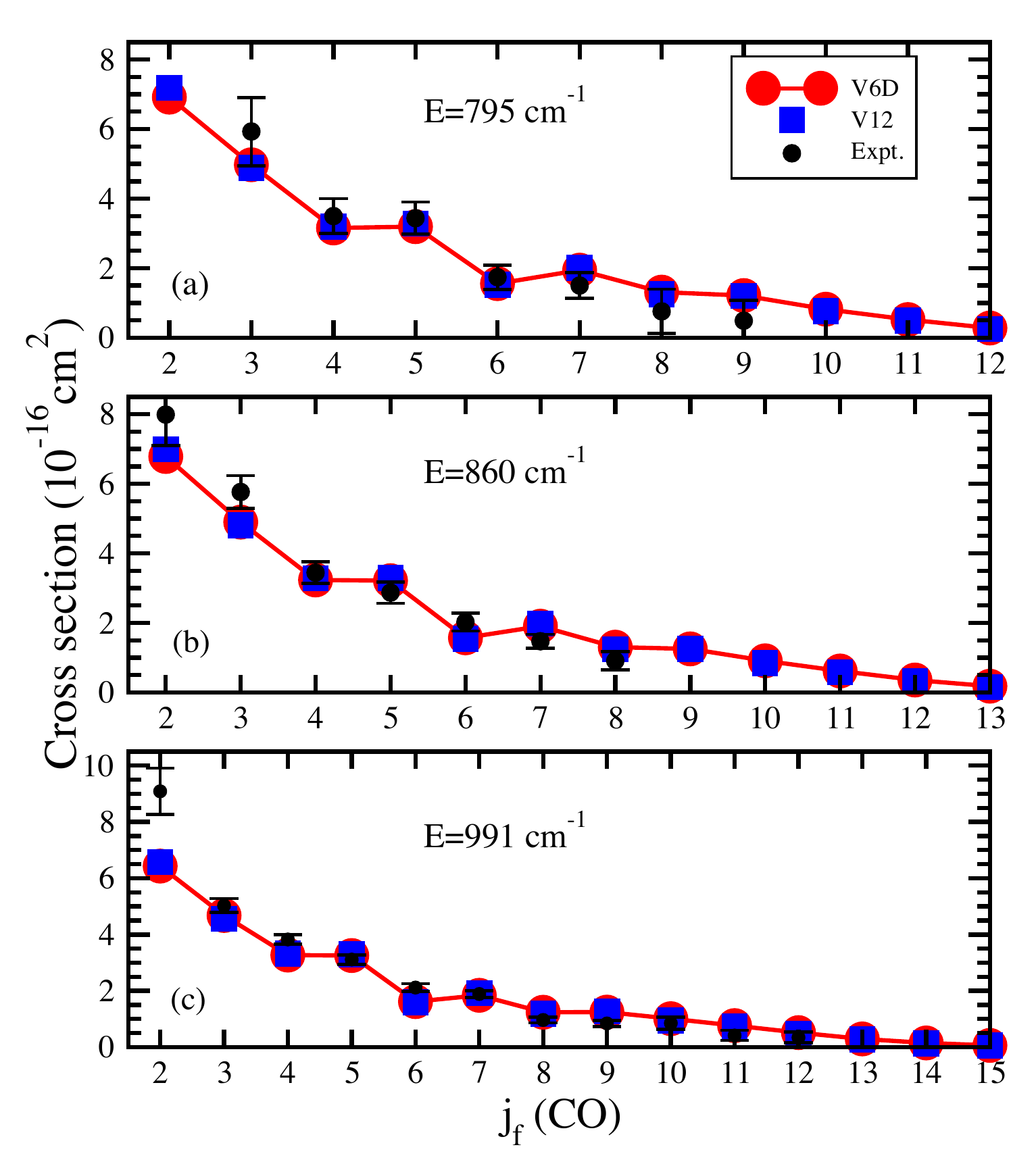}
\caption{
{\bf State-to-state cross sections for rotational excitation of CO($v_1=0,j_1=0,1$) by
 collisions with H$_2$.} Theoretical results of full-dimensional calculations on V6D and 4D rigid-rotor
calculations on V12 are compared with normalized experimental results \cite{ant00} for
collision energies of (a) 795 cm$^{-1}$, (b) 860 cm$^{-1}$, and (c) 991 cm$^{-1}$.
The calculations were performed for H$_2$($v_2=0,j_2=0$), but the experimental H$_2$ rotational
distribution was undetermined. The error bars correspond to twice the estimated standard deviation in the
weighted means of the measurements \cite{ant00}.}
\label{ant-exp}
\end{figure}

\begin{figure}[hbt]
\advance\leftskip -3.0cm
\includegraphics[scale=0.85]{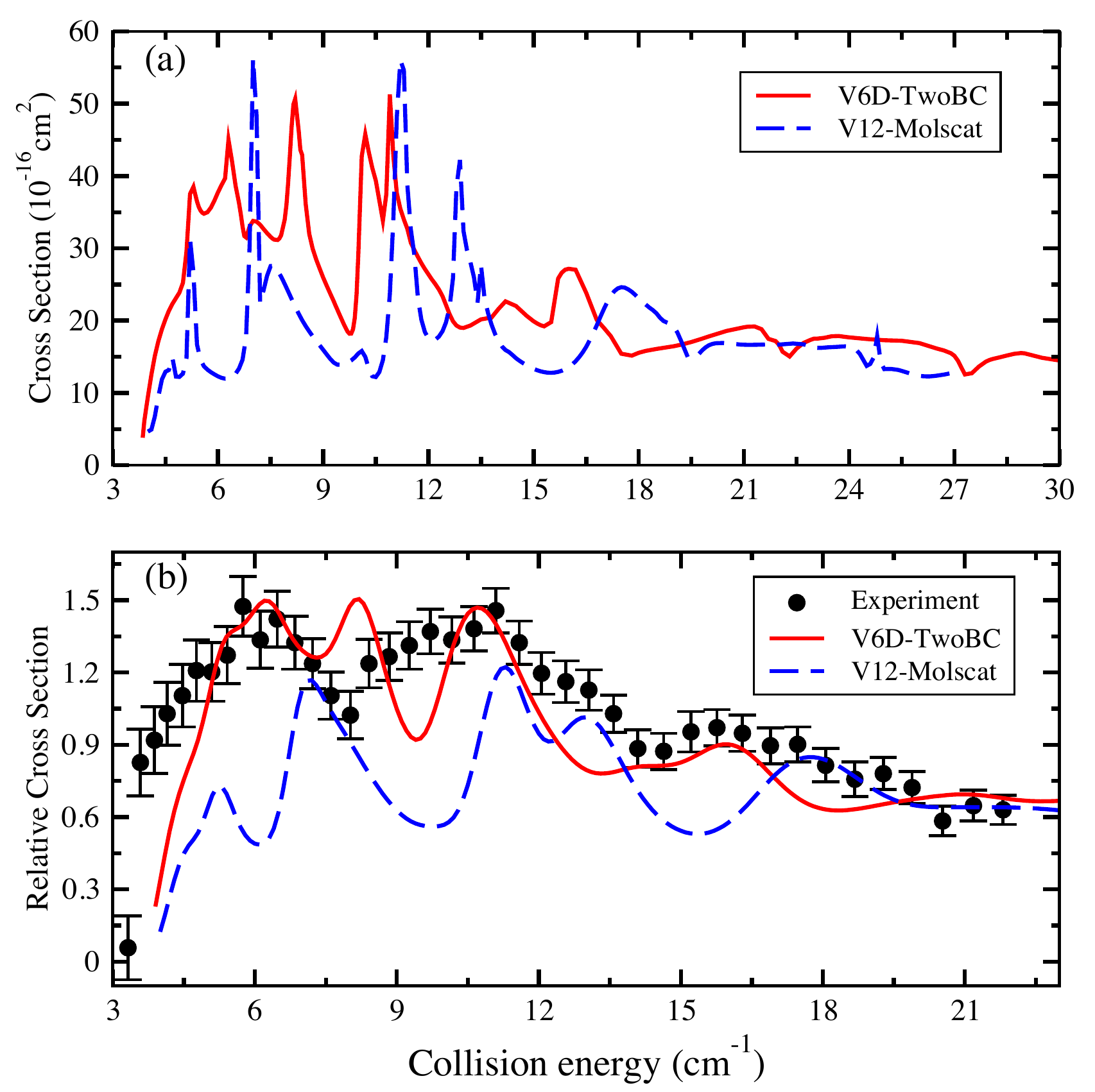}
\caption{ {\bf Low energy excitation cross sections.}
 $j_1=0 \rightarrow 1$ cross sections for CO($v_1=0$) due to
 collisions with H$_2$($v_2=0,j_2=0$) are shown as a function of collision energy.
 (a): computed cross sections using the 4D V12 and 6D V6D PESs.
 (b): computed cross sections convolved over the experimental beam energy spread
  (lines) compared to the relative experiment of Chefdeville {\it et al.} \cite{che12}
  (circles with error bars).  The error bars on the experimental cross sections of
  Chefdeville {\it et al.} represent the statistical uncertainty  at a 95\% confidence
 interval.
}
\label{gau-rigid}
\end{figure}

\begin{figure}[hbt]
\advance\leftskip -1.5cm
\includegraphics[scale=0.65]{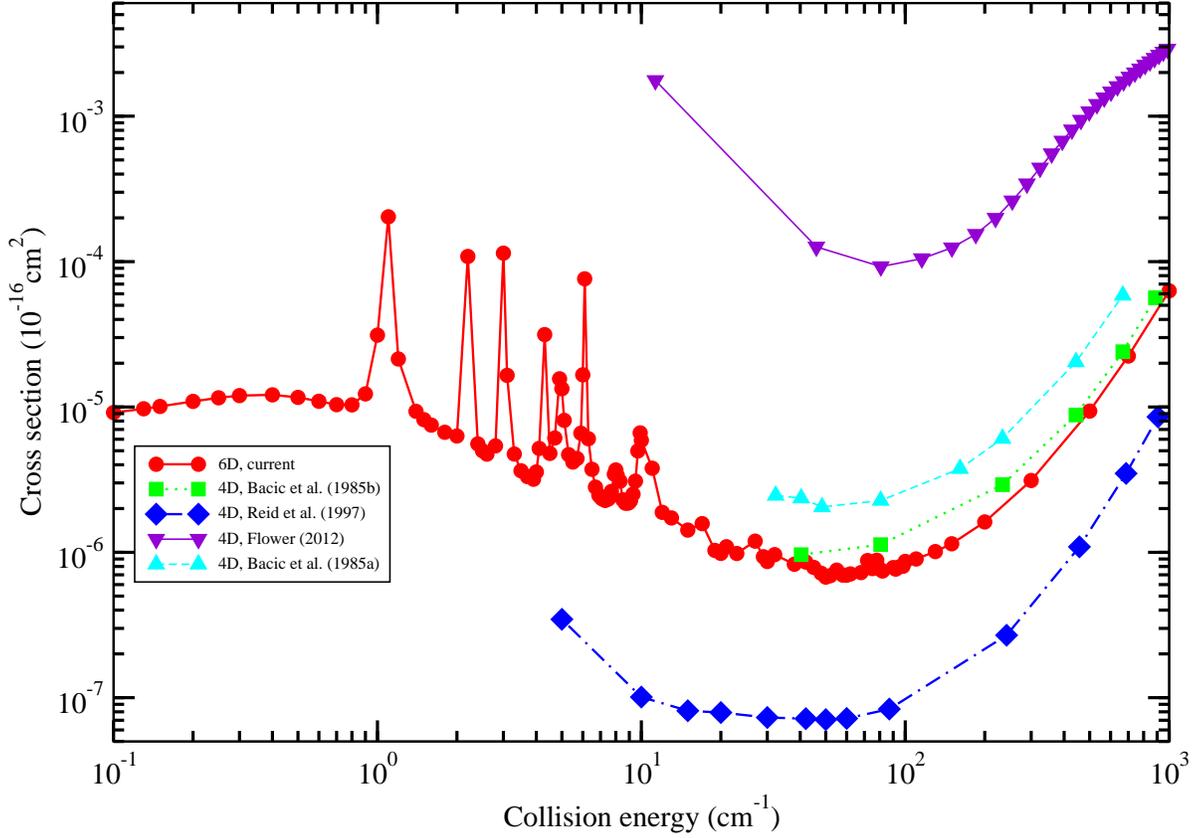}
\caption{
{\bf Total theoretical cross sections for the vibrational deexcitation of CO($v_1=1$) by para-H$_2$.}
Current 6D/CC results are compared to previous 4D calculations.
The 4D results of Ba\^ci\'c {\it et al.} \cite{bac85b} and Reid {\it et al.} \cite{rei97} do not
distinguish CO rotational states, while the 4D results of Ba\^ci\'c {\it et al.} \cite{bac85a}, 4D results of Flower \cite{flo12}, and the current 6D/CC
results are for initial $j_1=0$ summed over all final CO($v^\prime_1=0,j^\prime_1$).
In every case, the H$_2$ rovibrational state remains unchanged, $v_2,j_2 = v^\prime_2,
j^\prime_2=0,0$.
}
\label{total-cross}
\end{figure}

\begin{figure}[hbt]
\advance\leftskip -0.0cm
\includegraphics[scale=0.8]{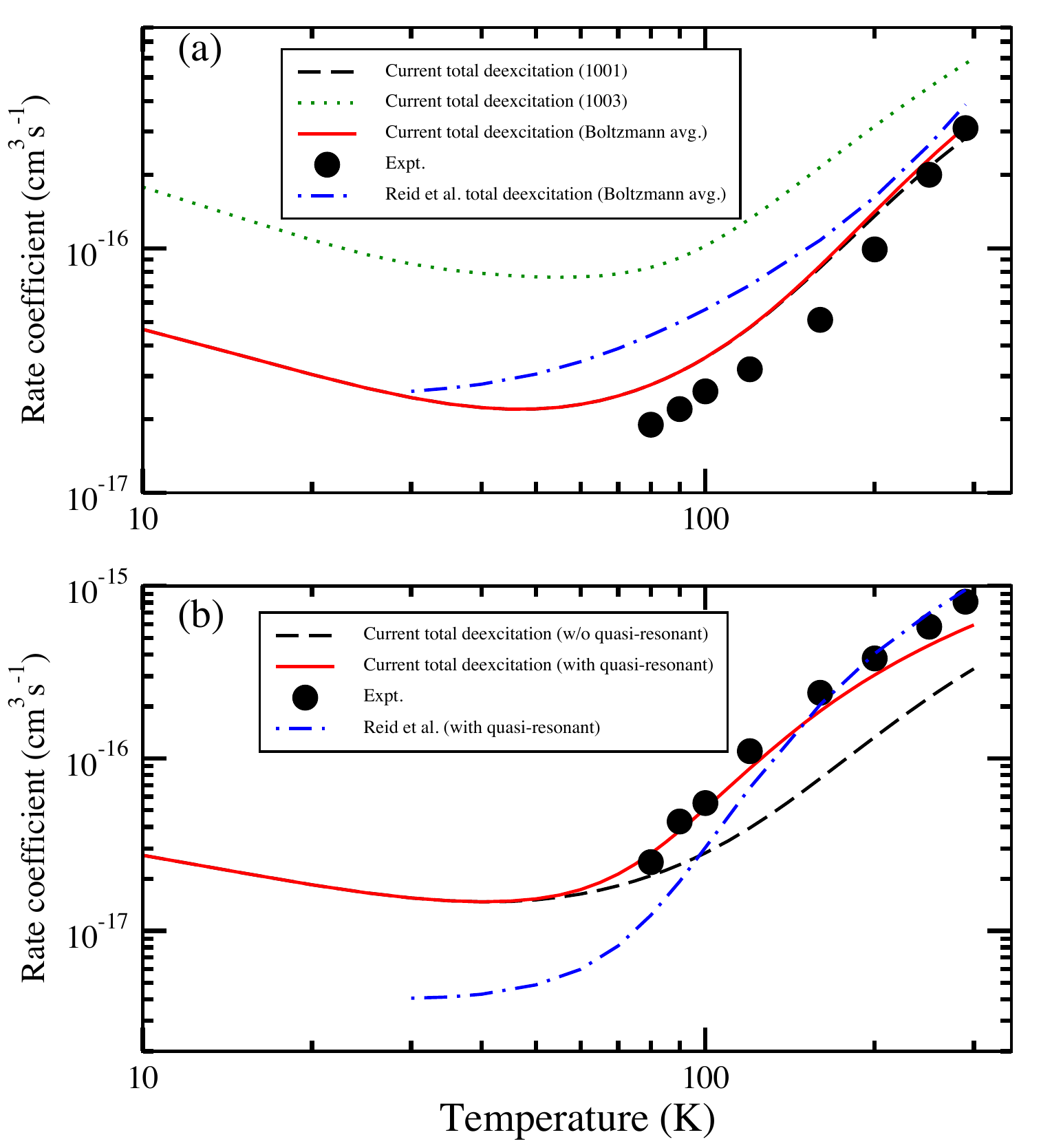}
\caption{
{\bf Rate coefficients for the vibrational deexcitation of CO($v_1=1$) due to H$_2$.}
Current 6D/CC (solid, dashed, and dotted lines) and Reid {\it et al.} \cite{rei97}
4D (dot-dashed line) calculations are compared to the total CO($v_1=1$) rovibrational
quenching experiment (symbols) \cite{and75,and76,wil93}. (a) Ortho-H$_2$ rate coefficients for
initial state-resolved, (100$j_2) \rightarrow (0j^\prime_10j_2^\prime$), summed
over $j^\prime_1$  and $j^\prime_2$ (dashed and dotted lines) and for a Boltzmann average of
initial H$_2$($j_2=1,3$) (solid and dot-dashed lines). (b) Para-H$_2$ rate coefficients for a Boltzmann average
of initial H$_2$($j_2=0,2$) with and without the quasi-resonant process (\ref{qr}). Note that the experimental
uncertainties are smaller than the symbol sizes.
}
\label{total-rate}
\end{figure}


\clearpage

\pagebreak
\begin{center}
\textbf{\LARGE  Supplementary   Information }
\end{center}
\setcounter{equation}{0}
\setcounter{figure}{0}
\setcounter{table}{0}
\setcounter{page}{1}
\makeatletter
\renewcommand{\theequation}{S\arabic{equation}}
\renewcommand{\thefigure}{S\arabic{figure}}
\renewcommand{\thetable}{S\arabic{table}}
\renewcommand{\bibnumfmt}[1]{[S#1]}
\renewcommand{\citenumfont}[1]{S#1}

\section*{Supplementary Figures }

\vskip 2.0cm

\begin{figure} [h]
\advance\leftskip -0.0cm
\includegraphics[scale=2.00, angle=0]{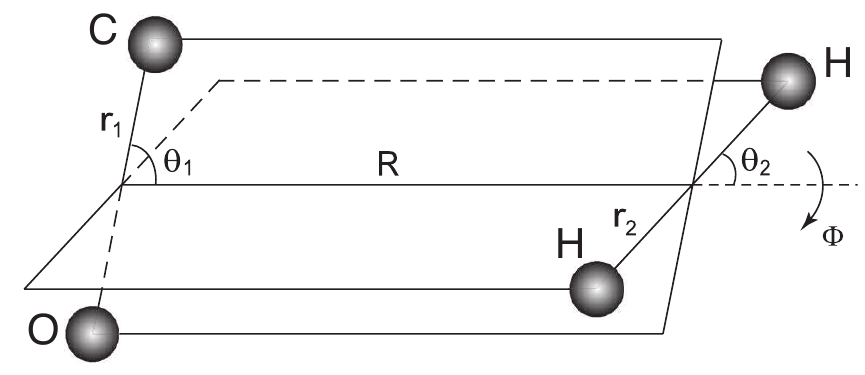}
\caption{
The six-dimensional Jacobi coordinates for CO-H$_2$.
$r_1$ and $r_2$ are bond lengths, $R$ the internuclear distance between CO and H$_2$
center of masses, $\theta_1$ and $\theta_2$ the angles between $\vec{R}$ and $\vec{r_1}$ and $\vec{r_2}$, 
and $\phi$ the dihedral or twist angle. 
}
\label{fig_jacobi}
\end{figure}

\clearpage

\begin{figure}
\advance\leftskip -3.1cm
\includegraphics[scale=0.85, angle=0]{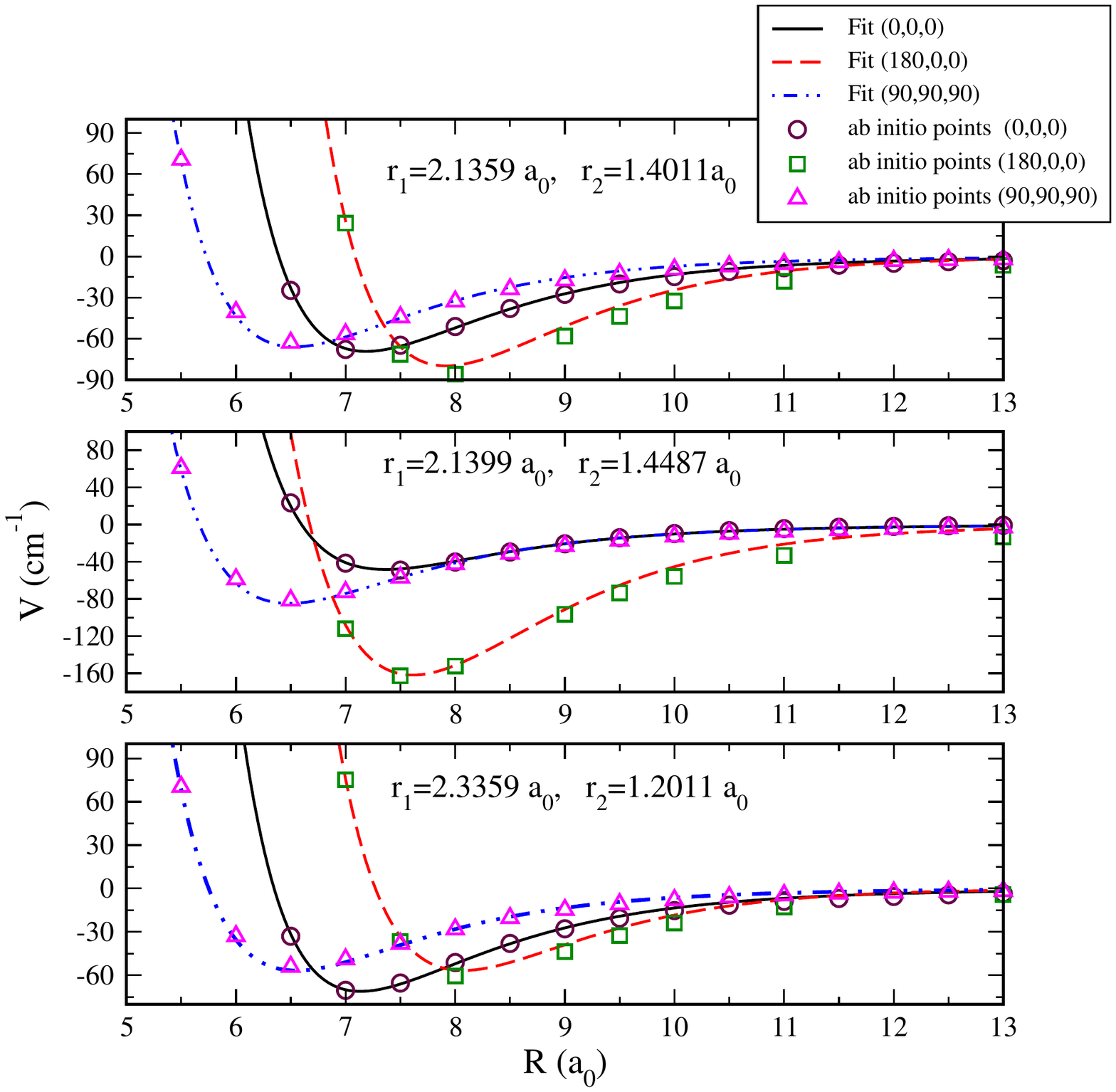}
\caption{Slices of the CO-H$_2$ interaction potential V6D. $R$ dependence of the interaction potential V6D for representative slices with bond lengths
fixed as indicated and
  ($\theta_1$, $\theta_2$, $\phi$)=($0^{\circ}$, $0^{\circ}$, $0^{\circ}$), ($180^{\circ}$, $0^{\circ}$,
 $0^{\circ}$), and ($90^{\circ}$, $90^{\circ}$, $90^{\circ}$). V6D fit (lines), computed ab initio
energy points (symbols). 
This figure may be compared to Fig. 2 of Ref. \cite{sjan98}.
}
\label{Vas}
\end{figure}

\clearpage

\begin{figure}
\advance\leftskip 0.0cm
\includegraphics[scale=0.85, angle=0]{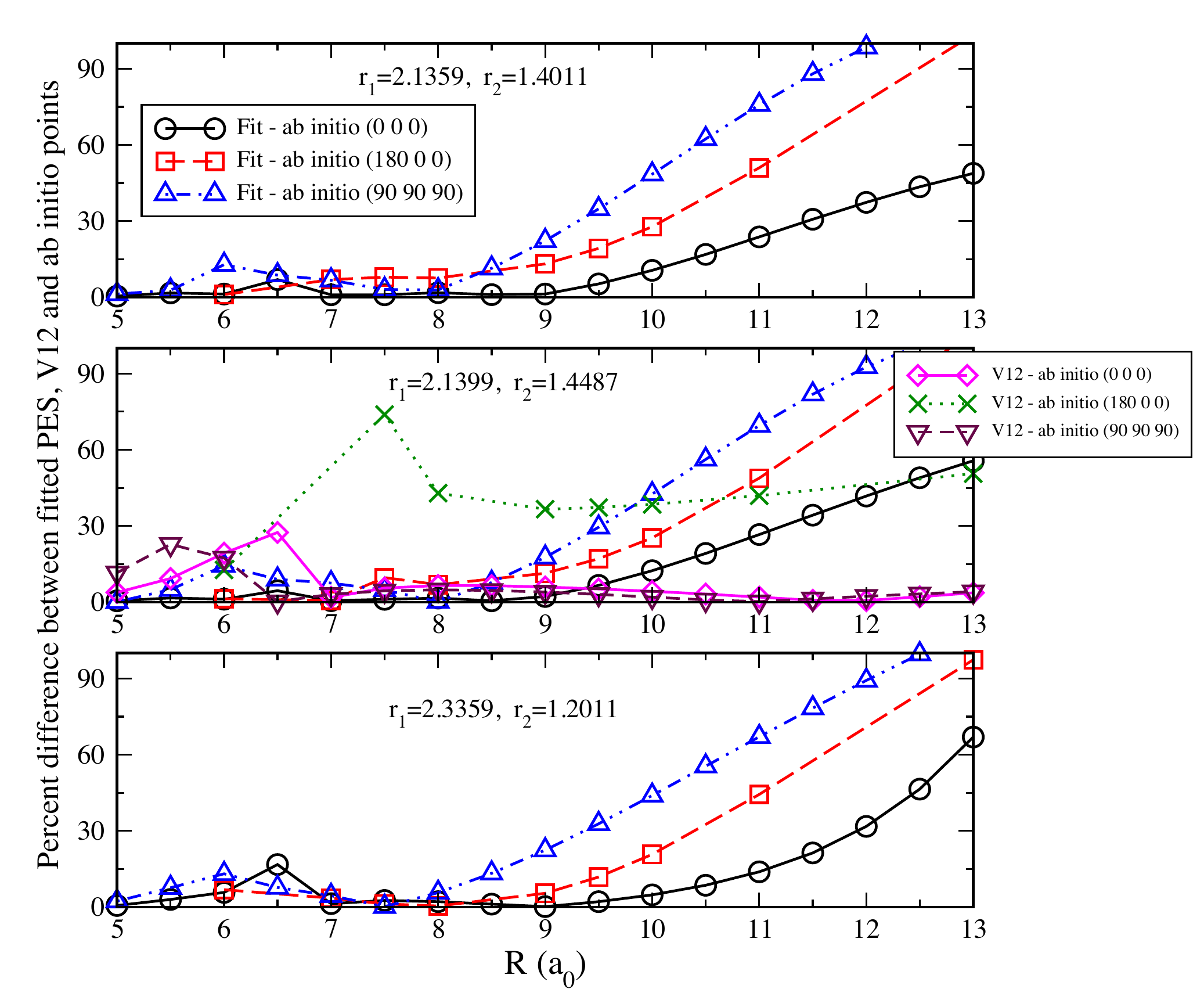}
\caption{Percent differences between CO-H$_2$ PESs. Percent
difference between the fitted PES and ab initio energy points
of the interaction potential V6D for representative slices with bond lengths
fixed as indicated and
  ($\theta_1$, $\theta_2$, $\phi$)=($0^{\circ}$, $0^{\circ}$, $0^{\circ}$), ($180^{\circ}$, $0^{\circ}$,
 $0^{\circ}$), and ($90^{\circ}$, $90^{\circ}$, $90^{\circ}$). The middle panel also displays the
 percent difference between the V12 PES and the current ab initio data. 
The percent differences only become large in the
asymptotic limit as the PES itself approaches zero. Also, displayed in the middle panel
is the difference between the V12 PES and the current
ab initio data. It is seen that there is good agreement, even in the long-range limit, except
for the $\theta_1=180^{\circ}$ configuration. This difference may partly be due to the fact that
V12 is a 4D PES obtained by averaging a 6D PES over the CO and H$_2$ ground state vibrational
wave functions. } 
\label{Vas1}
\end{figure}

\clearpage

\begin{figure}
\advance\leftskip -3.5cm
\includegraphics[scale=0.81, angle=0]{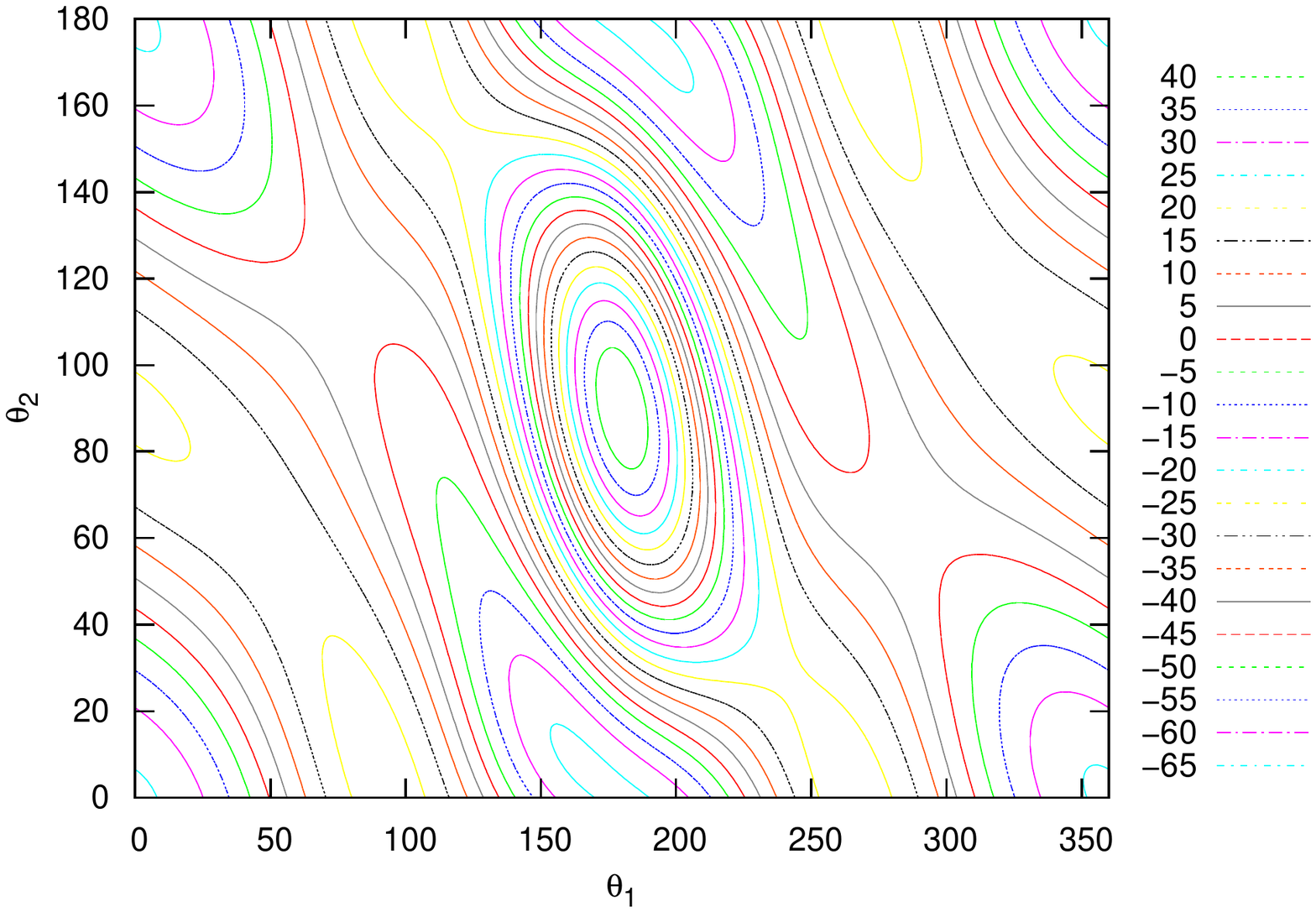}
\caption{Contour plot of the CO-H$_2$ V6D PES.  The contour plot is given as a function of $\theta_1$ and
  $\theta_2$.  $R=7.5$~a$_0$, $\phi=0^{\circ}$, $r_1$ = 2.1359~a$_0$,  $r_2$ = 1.4011~a$_0$.  
  The legend gives the contour energies in cm$^{-1}$.
  This figure may be compared to Fig. 1 of Ref. \cite{sjan98}.
}
\label{con_t1t2}
\end{figure}

\begin{figure}[h]
\advance\leftskip -2.5cm
\includegraphics[scale=0.70, angle=0]{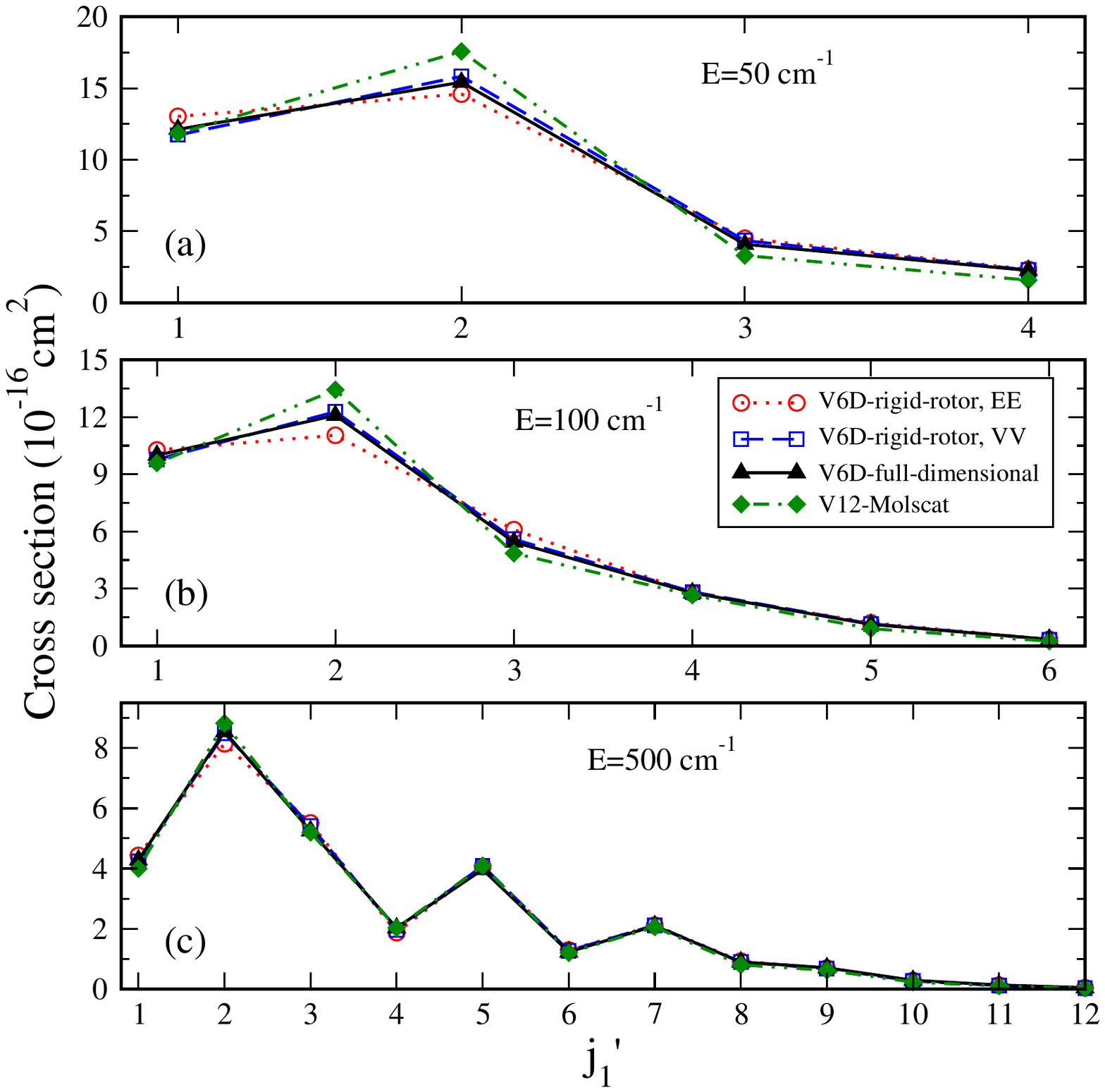}
\caption{CO rotational excitation cross sections due to H$_2$ collisions. Rigid-rotor approximation
 $j_1=0 \rightarrow j_1'$ excitation cross sections of CO($v_1=0,j_1$) by
 collisions with H$_2$($v_2=0,j_2=0$) using the V6D (EE and VV) and V12 potentials compared
 to the full-dimensional calculation. See Supplementary Note 1 for further details. Collision energy:
(a) 50 cm$^{-1}$, (b) 100 cm$^{-1}$, and (c) 500 cm$^{-1}$.
}
\label{e100-500}
\end{figure}

\clearpage

\begin{figure}
\advance\leftskip -2.5cm
\includegraphics[scale=0.80, angle=0]{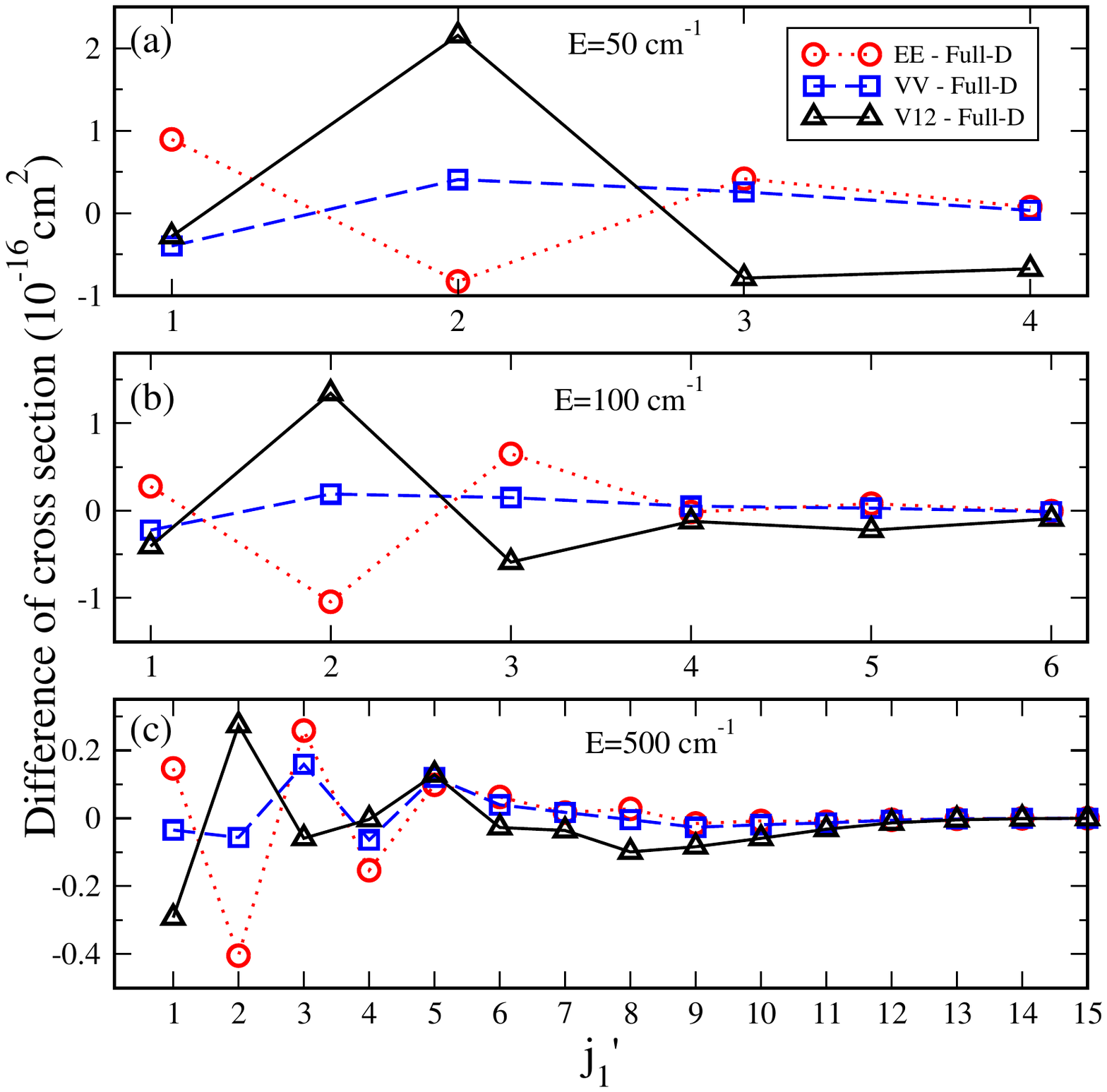}
\caption{CO rotational excitation cross section difference for H$_2$ collisions.
Difference of
 $j_1=0 \rightarrow j_1'$ excitation cross sections of CO($v_1=0,j_1$) by
 collisions with H$_2$($v_2=0,j_2=0$) using the V6D (EE and VV),  V12 potentials  and those
 of the full-dimensional calculation (Full-D, V6D). See Supplementary Note 1
 for further details. Collision energy:
(a) 50 cm$^{-1}$, (b) 100 cm$^{-1}$, and (c) 500 cm$^{-1}$.
}
\label{e50-500-df}
\end{figure}

\clearpage

\begin{figure}
\advance\leftskip -0.0cm
\includegraphics[scale=0.70, angle=0]{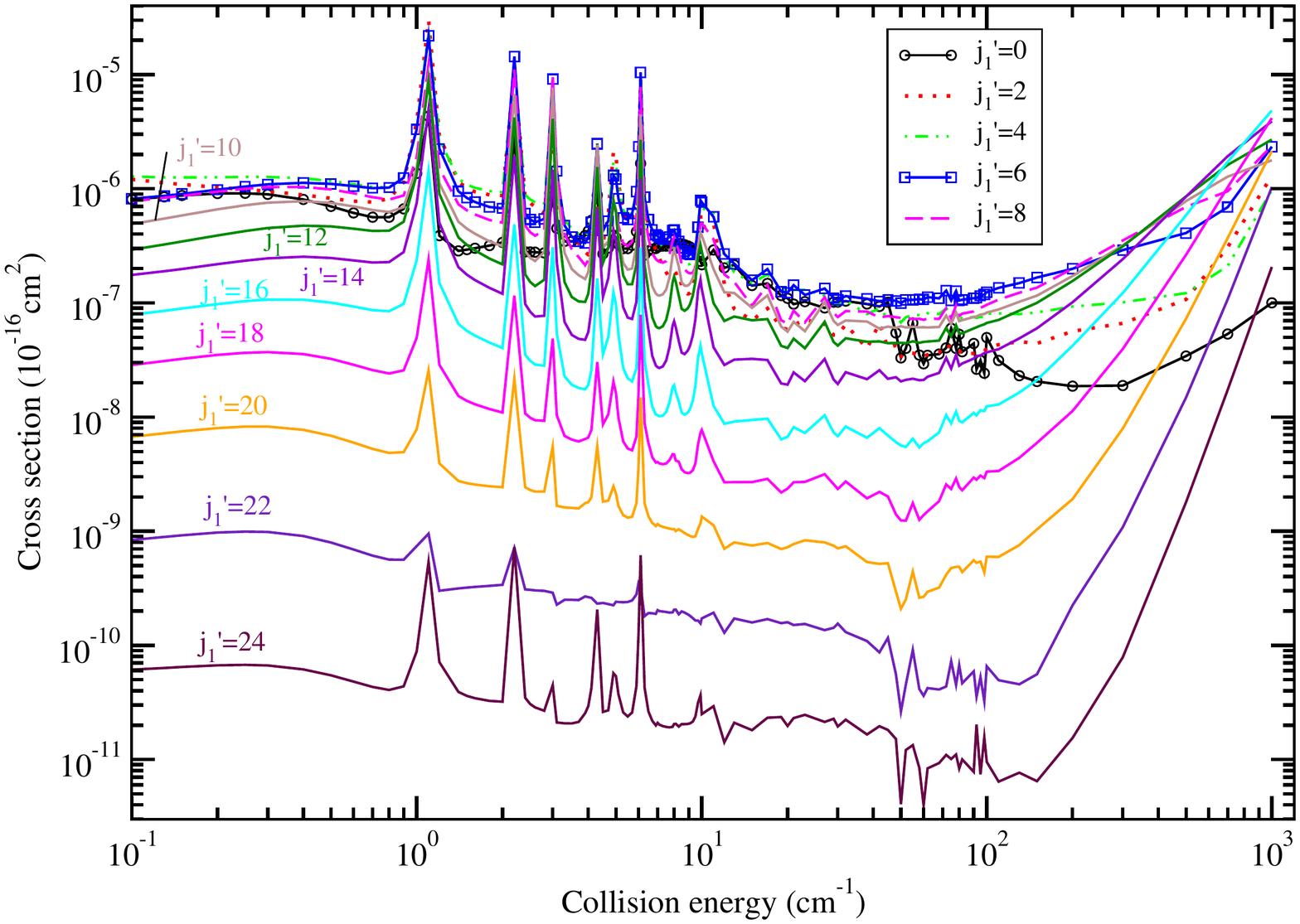}
\caption{CO--para-H$_2$ vibrational quenching state-to-state cross sections.
State-to-state cross sections for the vibrational quenching of CO from ($v_1=1, j_1=0$) to
($v_1^{\prime}=0, j_1^{\prime}$), $j_1^{\prime}$=0, 2, 4, $\cdots$, 24,
 due to para-H$_2$($v_2=0,j_2=0$) collisions, or in CMS notation (1000)$\rightarrow($0$j_1^\prime$00).
}
\label{st2st-p}
\end{figure}

\clearpage

\begin{figure}
\advance\leftskip -0.0cm
\includegraphics[scale=0.70, angle=0]{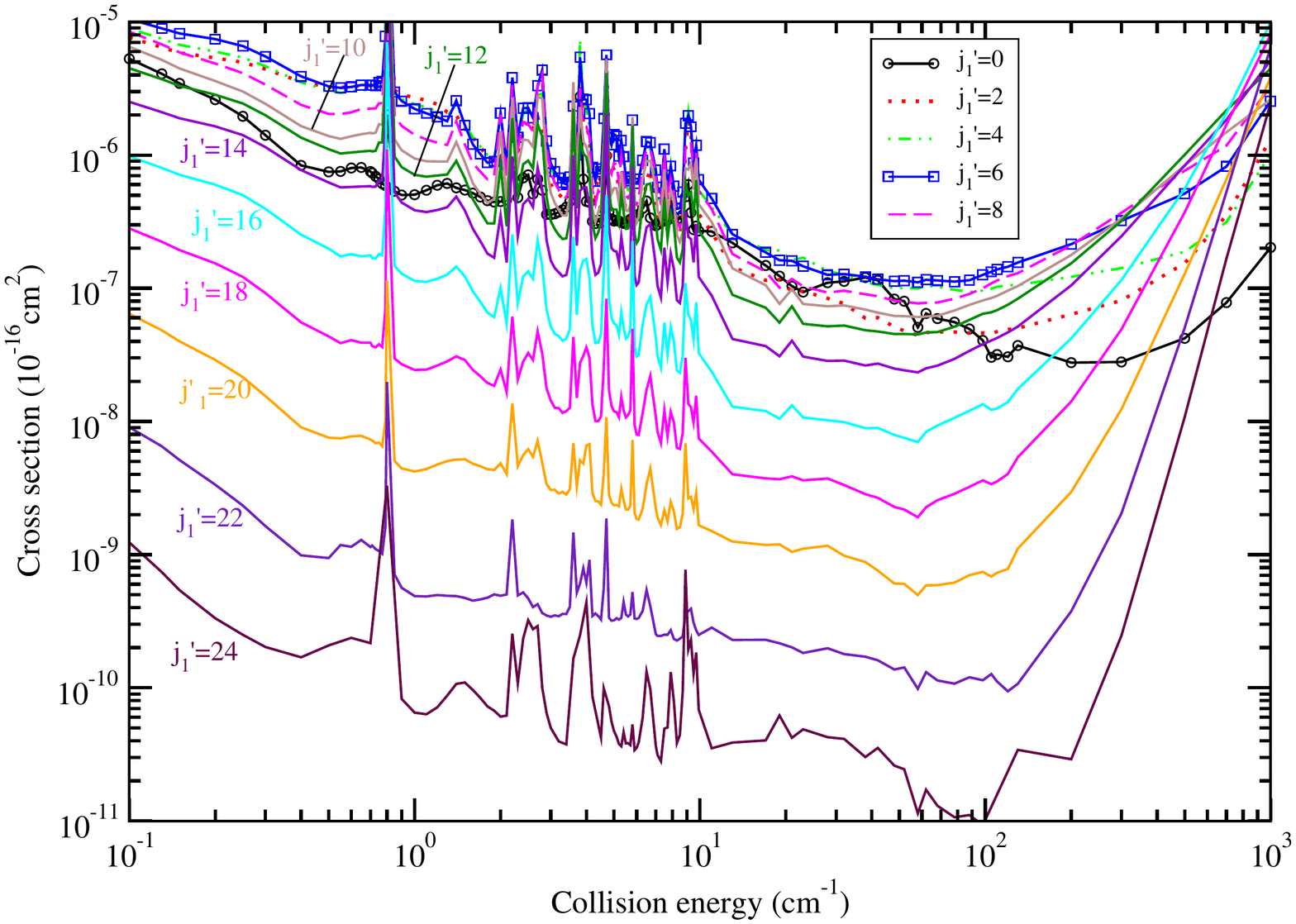}
\caption{CO--ortho-H$_2$ vibrational quenching state-to-state cross sections.
State-to-state cross sections for the vibrational quenching of CO from ($v_1=1, j_1=0$) to
($v_1^{\prime}=0, j_1^{\prime}$), $j_1^{\prime}$=0, 2, 4, $\cdots$, 24,
 by ortho-H$_2$($v_2=0,j_2=1$), or in CMS notation (1001)$\rightarrow($0$j_1^\prime$01).
}
\label{st2st-o}
\end{figure}

\clearpage

\begin{figure}
\advance\leftskip -1.5cm
\includegraphics[scale=0.70, angle=0]{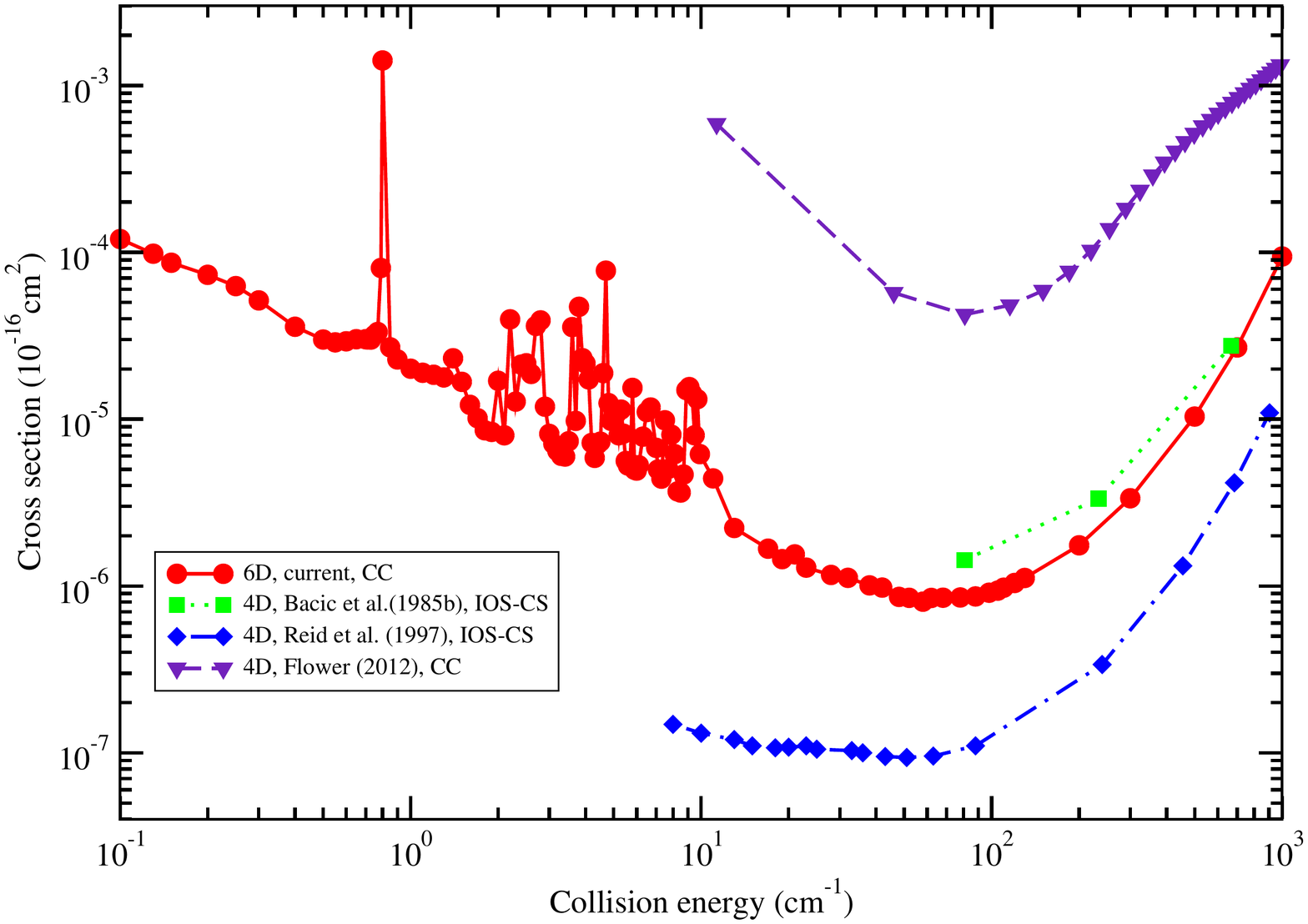}
\caption{Total CO--ortho-H$_2$ vibrational quenching cross sections.
Total cross sections for the vibrational quenching of CO($v_1=1$) + H$_2$($v_2=0,j_2=1$)
to CO($v_1^{\prime}=0$) + ortho-H$_2$($v_2^\prime=0,j_2^\prime$=1). Similar to
Fig. 4 of the main text, except for ortho-H$_2$.
}
\label{ortho00}
\end{figure}

\clearpage

\begin{figure}
\advance\leftskip -3.0cm
\includegraphics[scale=0.75, angle=0]{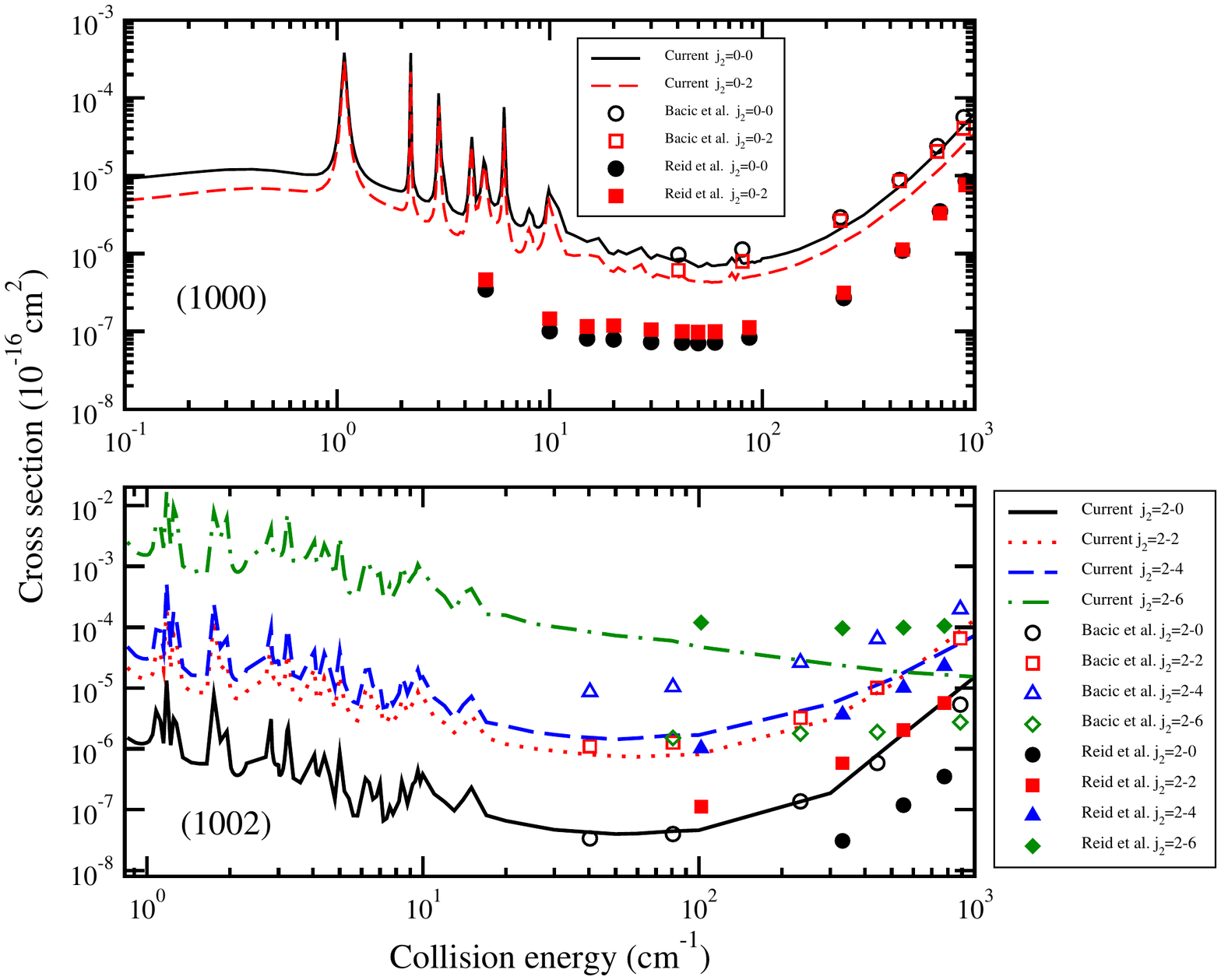}
\caption{Total CO--para-H$_2$ vibrational quenching cross sections for different H$_2$ states. 
Total cross sections for the vibrational quenching of CO from initial states (1000) (top panel) and (1002) (bottom panel)
to CO($v_1^{\prime}=0$) + para-H$_2$($j_2^\prime$=0, 2, 4, 6).
}
\label{para02}
\end{figure}

\clearpage

\begin{figure}
\advance\leftskip -2.0cm
\includegraphics[scale=0.80, angle=0]{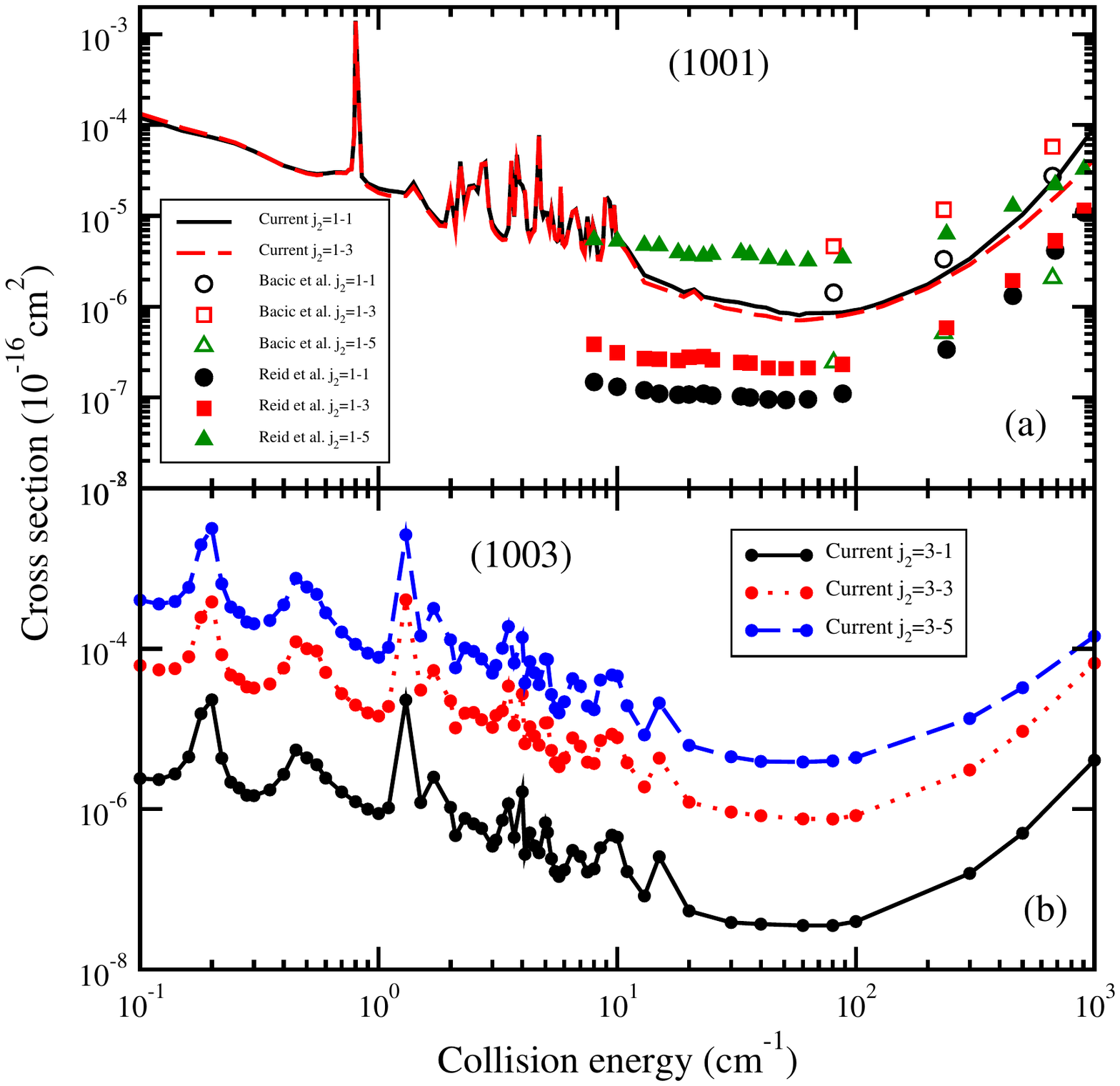}
\caption{Total CO--ortho-H$_2$ vibrational quenching cross sections for different H$_2$ states.
Total cross sections for the vibrational quenching of CO from initial states (1001) (top panel) and (1003) (bottom panel)
to CO($v_1^\prime$=0) + ortho-H$_2$($j_2^\prime$=1, 3, 5).
}
\label{ortho13}
\end{figure}

\clearpage

\section*{Supplementary Tables }

\begin{table}[h]
\begin{center}
\caption{CO-H$_2$ rotational excitation errors. RMS error (10$^{-16}$ cm$^2$) of the three rigid-rotor calculations and V6D in comparison with experiment 
\citep{sant00} at $E$ = 795, 860, and 991 cm$^{-1}$.}
\vskip 0.5cm
\begin{tabular}{c  @{\hspace{1.6cm}}  c @{\hspace{1.3cm}} c @{\hspace{1.3cm}} c @{\hspace{1.3cm}} c }
\tableline \tableline
    $E$ (cm$^{-1})$  &V6D-full &   V6D-EE  & V6D-VV   & V12    \\ [0.5ex]
\tableline
     795    & 0.549     & 0.546  & 0.552   & 0.556  \\   [1ex]
     860    & 0.571     & 0.598  & 0.572   &  0.551  \\   [1ex]
     991    & 0.946     & 0.989  & 0.946   &  0.892  \\   [1ex]
\tableline
\end{tabular}
\end{center}
\label{tabrms}
\end{table}

\clearpage

\begin{table}[h]
\begin{center}
\caption{Parameters used in the CO-H$_2$ scattering calculations.}
\vskip 0.5cm
\label{parameters}
\begin{tabular}{c c c c c c c c c  @{\hspace{0.8cm}} c}
\tableline \tableline 
   & Basis set\footnotemark[1] 
   & $N_{\theta_1}$ & $N_{\theta_2}$ &  $N_{\phi}$ & $N_{r_1}$ & $N_{r_2}$
    & $\lambda_1$ &  $\lambda_2$ & No. of channels 
   \\ [0.5ex]
\tableline \tableline
 4D rotational calculation & & & & & & & & &  \\ \tableline
TwoBC   & $j_1=30$, $j_2=2$ & 14 & 14  & 8 & 1 & 1 & 8 & 4 & 1344  \\   [1ex]
MOLSCAT &  $j_1=30$, $j_2=2$ & 14 & 14  & 8 &  &   & 8 & 4 & 1344  \\   [1ex] \tableline \tableline
 6D TwoBC calculation & & & & & & & & &  \\ \tableline 
 para-H$_2$-CO   & [(0,30;1,20)(0,2)] & 14 & 14  & 8 & 20 & 20 & 10 & 6 & 4258  \\   [1ex]
 para-H$_2$-CO   & [(0,22;1,20)(0,4)] & 14 & 14  & 8 & 20 & 20 & 10 & 6 & 7341  \\   [1ex]
 para-H$_2$-CO   & [(0,20;1,15)(0,6)] & 12 & 12  & 8 & 18 & 18 & 8  & 4 & 9832  \\   [1ex]
 ortho-H$_2$-CO   & [(0,30;1,20)(0,3)] & 14 & 14  & 8 & 20 & 20 & 10 & 6 & 5632  \\   [1ex]
 ortho-H$_2$-CO   & [(0,22;1,20)(0,5)] & 14 & 14  & 8 & 20 & 20 & 10 & 6 & 10251  \\   [1ex]
\tableline \tableline
\end{tabular}
\end{center}
 \footnotetext[1]{
Basis set [($v_1=0, j_{v_1=0}$; $v_1=1, j_{v_1=1}$)($v_2=0, j_{v_2=0}$)]
is presented by the maximum rotational quantum number $j_{v_1}$ and $j_{v_2}$
included in each relevant vibrational level $v_1$ and $v_2$ for CO and H$_2$,
respectively.
 }
\label{tab1}
\end{table}

\clearpage

\section*{Supplementary Discussion}

\subsection*{Supplementary Note 1:  Pure rotational scattering}

The V6D PES was adopted in CC scattering calculations using both TwoBC \cite{srom06} 
and MOLSCAT \cite{smolscat}.  Separate calculations were performed 
with the bond lengths of CO and H$_2$ fixed at: 
1) equilibrium bond lengths, $r_1$=2.1359~a$_0$ and $r_2$=1.4011~a$_0$, denoted as EE;
and 2) vibrationally-averaged bond lengths,
$r_1$=2.13992~a$_0$ and $r_2$=1.448736~a$_0$, denoted as VV. 
The 4-dimensional (4D) V12 potential of Jankowski {\it et al.} \cite{sjan13},     
which was constructed by averaging over their 6D potential energy data with CO($v_1=0$) and H$_2$($v_2=0$)
vibrational wave functions, 
was used in MOLSCAT calculations only.
For the rigid-rotor scattering calculations, TwoBC and MOLSCAT give 
almost identical CO state-to-state excitation cross section using the 4D version of V6D. As an example, 
Supplementary Fig.~\ref{e100-500} displays the CO state-to-state excitation cross sections from $j_1=0$
for collision energies of 50, 100, and 500 cm$^{-1}$ due to H$_2$($v_2=0,j_2=0$) using the V12 potential
with MOLSCAT
 and the two 4D (EE and VV) versions of V6D with TwoBC. 
 Supplementary Fig.~\ref{e50-500-df} gives the differences in these cross sections compared to
V6D (full-dimensional) results.
There is generally good
agreement among all approaches with a small difference depending on which CO and H$_2$ 
bond-lengths are adopted. Comparison is also made to the full 6D calculation which agrees best
with the rigid-rotor VV results. This confirms both the validity of the rigid-rotor approximation
and the suggestion of Jankowski and Szalewicz \cite{sjan05}  that PESs constructed with
vibrationally-averaged bond lengths are preferred for scattering calculations. Further, the discrepancy
in the rigid-rotor calculations and the 6D results are seen to increase, at least for 
low $j^\prime_1$, with decreasing collision
energy supporting the claim in the main text that accurate prediction 
of low-energy dynamics requires full-dimensionality.

Antonova {\it et al.} \cite{sant00} measured the relative state-to-state
rotationally inelastic cross sections for excitation of CO by H$_2$ in
a crossed molecular beam experiment at collision energies of 795, 860,
and 991 cm$^{-1}$ as discussed in the main text. Comparison between
the current calculated cross sections 
 and the measurements give a
root mean square (RMS) error computed by, 
\begin{equation}
 \textrm{RMS} = \sqrt{\frac{\sum\limits_{i}^{N} (\sigma_{\textrm{expt}}-
 \sigma_{\textrm{theo}})^2 } {N} },  
 \label{eq_rms}
\end{equation}
where $\sigma$ is the cross section and $N$ the number of data points.  The RMS errors for 
 the 4D version of V6D with EE and VV bond lengths, the 4D V12 PES, and the
V6D PES  
are shown in Supplementary Table~1 with the cross sections for the latter two
shown in Fig. 2 of the main text.
The current V6D results for the final CO $j_1^{\prime}$ distributions (Fig. 1) are seen to
be in excellent agreement with experiment,
validating the accuracy of the fitted potential as well as the 
present scattering calculations.

Chefdeville {\it et al.} \cite{sche12} reported crossed-beam experiments and 
quantum-mechanical calculations performed for the 
CO($j_1$=0) + H$_2$($j_2$=0) $\rightarrow$ CO($j_1^{\prime}$=1) + H$_2$($j_2^{\prime}$=0) process. 
The 4D V04 PES of Jankowski and Szalewicz \citep{sjan05} was used in their scattering calculations. 
To compare with the measurements,
the computed cross sections were convolved with the experimental kinetic energy distribution according to,
\begin{equation}
  \bar{\sigma}(E) = \frac{1}{\sqrt{2\pi}\delta}\int_0^{\infty} \sigma(E')
   \exp \left[-\frac{(E-E')^2}{2\delta^2} \right]dE',
 \label{equ_gau}
\end{equation} 
where $\delta$ is given by \cite{sche12}, 
\begin{equation}
\delta=\sqrt{2}(0.077+0.051E-3.6\times 10^{-6} E^2). 
\end{equation} 
The detailed cross sections and discussion are presented in the main text.

\subsection*{Supplementary Note 2:  Full-dimensional Scattering calculations}

\subsubsection*{Quenching cross sections}

We have performed full-dimensional calculations of the state-to-state cross
sections for initial CMSs (100$j_2$), for para-H$_2$ ($j_2$=0 and 2), and ortho-H$_2$ ($j_2$=1 and 3),
for collision energies ranging from 0.1 to 1000 cm$^{-1}$. 
As examples, in Supplementary Figs.~\ref{st2st-p} and \ref{st2st-o} the state-to-state quenching
cross sections for final states (0$j_1^{\prime}$0$j_2$), 
$j_1^{\prime}$=0, 2, 4, $\cdots$, 24 are shown for para- and
ortho-H$_2$, respectively, where $j_2=j_2^\prime$, i.e., elastic in H$_2$. 
The cross sections have similar behaviour and display resonances 
in the intermediate energy region primarily due to  
quasibound states supported by the van der Waals well of the interaction potential 
of the initial channel. Resonances and oscillatory structures, which will be analyzed
further in a future publication,  appear up
to about 100 cm$^{-1}$, roughly the depth of the van der Waals minimum.
Small $|\Delta j_1|=|j_1^{\prime}-j_1|$  transitions dominate the 
quenching; the cross sections generally decreasing
with increasing $j_1^{\prime}$ with that for $j_1^{\prime}=24$ being the smallest.

The current 6D/CC state-to-state cross sections are summed over $j_1'$ (for $v_1'=0$) to compare to the total deexcitation
4D cross sections given by Ba\^ci\'c {\it et al.} \cite{sbac85a,sbac85b}, Reid {\it et al.} \cite{srei97}, and
Flower \cite{sflo12} as shown for para-H$_2$ in Fig.~4 of the main text and here for ortho-H$_2$ in Supplementary Fig.~\ref{ortho00}.
As discussed in the main text there is a two order of magnitude dispersion between the various calculations
which may partly be explained by the differing treatments of angular momentum coupling.
In the infinite order sudden (IOS) approximation, the internal
rotation is neglected (effectively setting $j=0$), while the coupled-states (CS) approach,
which is a centrifugal sudden approximation, allows for the dynamical treatment of rotational
transitions. Ba\^ci\'c {\it et al.} \cite{sbac85a} performed both CS and IOS
calculations, but with $j_2=0$ (i.e., only para-H$_2$), while Ba\^ci\'c {\it et al.}
\cite{sbac85b} and Reid {\it et al.} \cite{srei97} used a mixed IOS-CS formalism
in which the rotational motion of CO was neglected, but that of H$_2$ retained,
allowing for significant reduction of basis sets.
The cross sections from Flower \cite{sflo12} are likely too large due
to an insufficient CO basis, an effect studied by Ba\^ci\'c {\it et al.} \cite{sbac85a}.
In fact, using Flower's basis set in CC calculations on V6D, we obtained
significantly larger cross sections than our converged results, verifying the primary cause for the
discrepancy. However, in most cases differences are related to the adopted PESs and dimensionality
of the scattering approaches.

\subsubsection*{Quenching rate coefficients}

In the calculations of Ba\^ci\'c {\it et al.} \cite{sbac85a,sbac85b} and Reid {\it et al.} \cite{srei97}, 
the total quenching cross section from state ($v_1=1, j_2$) was obtained
by summing the state-to-state quenching cross sections over the final rotational state of 
H$_2$, $j_2^{\prime}$, 

\begin{equation}
\sigma(v_1=1, j_2 \rightarrow v_1^{\prime}=0)(E) 
    = \sum_{{j_2}^{\prime}} \sigma(v_1=1, j_2 
        \rightarrow v_1^{\prime}=0, j_2^{\prime})(E),
\end{equation}
allowing for H$_2$ rotational inelastic transitions (with
$v_2=v_2^\prime =0$ and no information on $j_1$ or $j_1^\prime$ as the IOS approximation
was adopted for CO in both previous calculations.)
By thermally averaging the cross sections over the collision energy, the
quenching rate coefficients from $v_1=1, j_2$ were obtained. 

The total thermal rate coefficients for CO $v_1=1 \rightarrow v_1^\prime=0$  vibrational quenching
were obtained by 
summing $k(v_1=1, j_2 \rightarrow v_1^{\prime}=0)(T)$
over all initial H$_2$ rotational states $j_2$ weighted by
their populations assuming a Boltzmann distribution at $T$.
As observed by Millikan and Osburg \cite{smil64},  para-H$_2$ is a more efficient
collision partner than ortho-H$_2$ for vibrational quenching of CO due to  
the quasi-resonant transition: \\
CO($v_1=1$) + H$_2$($v_2=0, j_2=2$) $\rightarrow$ 
CO($v_1^{\prime}=0$) + H$_2$($v_2^{\prime}=0, j_2^{\prime}=6$).
Guided by the measurements, this quasiresonant process was also theoretically investigated by Ba\^ci\'c {\it et al.} \cite{sbac85b} 
and Reid {\it et al.} \cite{srei97} requiring cross sections for rotationally excited H$_2$ with
inelastic H$_2$ transitions. Supplementary Figs.~\ref{para02} and \ref{ortho13} compare the $j_2$ state-to-state cross
sections from the earlier work with the current results. As shown in the top panel of Supplementary Fig.~\ref{para02},
the current 6D/CC results are in reasonable agreement with the 4D/CS-IOS calculations of Bac\^i\'c {\it et al.}
for the 1000$\rightarrow$0$j_1^\prime$00 and  1000$\rightarrow$0$j_1^\prime$02 cases,
but both results are nearly an order of magnitude larger than obtained by Reid {\it et al.} In the
bottom panel of Supplementary Fig.~\ref{para02}, a large scatter in the results is evident for initial state (1002), but generally
the cross sections increase with increasing $\Delta j_2 = j_2^\prime - j_2$. The important exception is
that contrary to both the current calculations and the 4D/CS-IOS results for Reid {\it et al.}, Ba\^ci\'c
{\it et al.} find much smaller cross sections for the quasi-resonant process. The V6D/CC results shown
in that figure for the quasi-resonant transition were the most computationally time-intensive 
due to the larger size of the H$_2$ basis 
and may represent a lower limit for the cross section at the highest energies.

Supplementary Fig.~\ref{ortho13} gives a similar comparison for ortho-H$_2$ collisions though a quasi-resonant
process is not operable due to much larger asymptotic energy differences. Otherwise, the trends
are very similar to those noted for para-H$_2$ collisions, though the current V6D/CC results fall somewhat
between the earlier two 4D calculations. Note that we did not consider the H$_2$ inelastic channel
$j_2=1\rightarrow 5$ as Bac\^i\'c {\it et al.} \cite{sbac85b} find the cross sections to be more than an order of magnitude smaller
than the $j_2=1\rightarrow 3$ transition. This is in contrast to Reid {\it et al.} who find the  $j_2=1\rightarrow 5$ transition to
dominate the quenching of (1001). As a consequence, the thermally-averaged rate coefficients for
the quenching by ortho-H$_2$ obtained by Reid {\it et al.}, shown in Fig.~5a of the main text, are significantly
larger than the V6D/CC computation and the experiment. These rate coefficients also include quenching from the
initial state (1003), for which the cross sections are shown in the bottom panel of Supplementary Fig.~\ref{ortho13},
though Reid {\it et al.} do not tabulate them. 
Because of the complexity
of the system and differences in the PESs and scattering treatments, it is difficult to draw 
conclusions from one-to-one comparisons of the various calculations. However, the main
deficiency in the IOS approach is that it does not resolve the final CO($j_1^\prime$) channels
which can play a significant role in the cross sections as illustrated in 
Supplementary Figs.~\ref{st2st-p} and \ref{st2st-o}. 


\section*{Supplementary References}

\end{document}